\RequirePackage[2018-12-01]{latexrelease}
\documentclass[showpacs,showkeys,11pt,
preprint,preprintnumbers,nofootinbib,
groupedaddress,superscriptaddress,amsmath,amssymb]{revtex4}
\usepackage{cancel}
\usepackage{xcolor}
\usepackage{tabularx}
\usepackage{amsfonts}
\usepackage{amsmath}
\usepackage{graphicx,subfigure}
\usepackage{caption}
\usepackage[export]{adjustbox}% http://ctan.org/pkg/adjustbox
\usepackage{verbatim} 
\usepackage{epsfig}
\usepackage{url}
\usepackage{multirow}
\usepackage{hhline}
\usepackage{feynmp}
\usepackage{booktabs}
\usepackage{csquotes}

\newcommand {\be}{\begin{equation}}
\newcommand {\ee}{\end{equation}}
\newcommand {\ba}{\begin{eqnarray}}
\newcommand {\ea}{\end{eqnarray}}

\begin{document}
\title{Leptoquark Searches at TeV Scale Using Neural Networks at the Hadron Collider}

%%%%%%%%%%%%%%%%%%%%%%%%%%%%%%%%%%%%%%%%%%%%%%%%%%%%%%%%%%%%%%%%%%%%%%%%%%%%%%%%
\author{Ijaz Ahmed}
\email{ijaz.ahmed@fuuast.edu.pk}
\affiliation{Federal Urdu University of Arts, Science and Technology, Islamabad, Pakistan}
\author{Usman Ahmad}
\email{usmanahmed1661@gmail.com}
\affiliation{Riphah International University, Islamabad, Pakistan}
\author{Jamil Muhammad}
\email{mjamil@konkuk.ac.kr}
\affiliation{Sang-Ho College \& Department of Physics, Konkuk University, Seoul 05029, South Korea}
\author{Saba Shafaq}
\email{saba.shafaq@iiu.edu.pk}
\affiliation{International Islamic University, Islamabad, Pakistan}
%%%%%%%%%%%%%%%%%%%%%%%%%%%%%%%%%%%%%%%%%%%%%%%%%%%%%%%%%%%%%%%%%%%%%%%%%%%%%%%%
\date{\today}

\begin{abstract}

%Leptoquarks (LQs) may be responsible for recently reported discrepancies from the Standard Model prediction in B meson decays and anomalous magnetic moment of the muon. In the Standard Model (SM), leptoquarks are hypothetical particles that link the quark and lepton sectors. The phenomenology of leptoquarks hence becomes quite intriguing and has been thoroughly researched recently. Typically, the theories that account for the anomalies include LQs that pair to second- and third-generation leptons and quarks. This provides more justification for looking for such LQs( frequently referred to as second and third-generation LQs ) at the LHC.\\ 
%This paper presents a search for pairs of leptoquarks formed in proton-proton collisions with a centre-of-mass energy  $\sqrt{s}=14$~TeV.

Several discrepancies in the decay of B-meson decay have drawn a lot of interest in leptoquarks (LQ), making them an exciting discovery. The current investigation aims to discover the pair-production of leptoquarks that links strongly to the third generation of quarks and leptons at the center of mass energy $\sqrt{s}$=14 TeV via proton-proton collisions at the Large Hadron Collider (LHC). Based on the lepton-quark coupling parameters and branching fractions, we separated our search into various benchmark points. The leading order (LO) signals and background processes are generated, while parton showering and hadronization are also performed to simulate the detector effects. \\
This article examines the hadronic and semileptonic decay modes of Leptoquark at three different mass values 1350, 1500 and 1650 GeV. 
  We analyze the signal significance at three integrated luminosities 300 $fb^{-1}$, 1000 $fb^{-1}$, and 3000 $fb^{-1}$. The comprehensive analysis of hadronic decay as well as semileptonic channels provides a significance of 4.2$\sigma$ at a typical point of 1500 GeV and 1000 $fb^{-1}$. We apply six distinct classifiers available in multivariate analysis (MVA) algorithms at the same mass and luminosity to increase sensitivity.  The hadronic channel significance is increased to 6.2$\sigma$ using the MVA method, indicating a significant improvement.  The semileptonic decay is consistently more significant than the hadronic mode across all benchmarks. Our results demonstrate how well machine learning methods might enhance the possibility of discovering scalar leptoquarks during present and next LHC operations.%A detailed comparison of the final results with the conventional methods is also presented in which the Madanalysis and MVA show higher significance at lower masses of scalar leptoquark at $\mathcal{L}_{int} = 1000fb^{-1}$ and $\mathcal{L}_{int}=3000fb^{-1}$.}
\end{abstract}

\maketitle

\section{Introduction}
%So far, the precision with which the Standard Model (SM) predictions have been confirmed is astonishing. However, certain enduring discrepancies in rare B-meson decays discovered across multiple distinct investigations suggest the existence of the novel physics. For instance, the BaBar collaboration revealed an important surplus in the $R_{D}^{*}$ observables for the first time in 2012 \cite{lees2012evidence,bhattacharya2015simultaneous}. Among the several extensions of SM, the simplest one is a scenario which, involves a tree-level exchange of Leptoquarks (LQs). These are the particles that may convert a lepton into a quark and vice versa, and establish a link between the lepton and quark sectors in the SM \cite{pati1974lepton, georgi1974unity,dimopoulos1979mass,dimopoulos1980technicoloured, eichten1980dynamical, angelopoulos1987search, buchmuller1986constraints}, which have possible comparable structures.\\ %Many extensions of the standard model (SM) of particle physics, such as those based on the concepts of grand unification \cite{georgi1974unity}, technicolor \cite{hill2003strong}, or compositeness \cite{d1992preons}, anticipate them.

Up to a great degree of testable accuracy, the Standard Model (SM) has already sufficiently described the color and electroweak sectors. The predicted particle spectrum of the SM has also been completed with the unveiling of the 125 GeV Higgs boson at the Large Hadron Collider (LHC) \cite{atlas2012observation,cms2012observation}. Despite this, several theoretical problems and experimental findings cannot be explained by the SM, suggesting the existence of some New Physics (NP) that has not yet been investigated. Gauge coupling unification, for instance, suggests a more substantial theory that corresponds to a single gauge group. The effective low-energy shape of the SM gauge group, $SU(3)_C \times SU(2)_L \times U(1)_Y$, emerges using a certain symmetry-breaking chain. Such Grand Unified Theories (GUT) include SU (4) \cite{pati1974lepton}, SU (5) \cite{georgi1974unity}, SO (10) \cite{georgi1975state,fritzsch1975unified}, $E_6$ \cite{kang2008theory,hati2016overline}, and others. It should be noted that quarks and leptons can pair directly at the tree level within a GUT structure via the hypothetical Leptoquark (LQ) mediator \cite{dorvsner2016physics,davidson1994model,hewett1997much,nath2007proton}. Although LQs can be either scalar or vector in a local quantum field theory, scalar LQs are more helpful for examining the loop-induced Beyond Standard Model (BSM) implications \cite{bl1997leptoquark,fajfer2016vector,barbieri2016anomalies}. LQs are important from a number of phenomenological perspectives.  In particular, several types of deviations of the B-meson \cite{dorvsner2013minimally,gripaios2015composite,bevcirevic2015lepton,bevcirevic2016leptoquark,crivellin2017simultaneous,cline2018b,di2017scale,aydemir2020addressing,crivellin2020flavor,asadi2023wrinkles} can be explained by an extension of the SM through the LQ. Leptoquarks could also be important for the generation of scalar particles in the LHC \cite{bhaskar2020enhancing,da2021enhancement,agrawal2000leptoquark,enkhbat2014scalar}, as well as dark matter physics \cite{choi2018lepto,mohamadnejad2019accidental}.
For an improvement over the proton lifetime limitations, the most basic GUT extensions require a heavy LQ \cite{abe2014search,dorvsner2012heavy}, but these are not feasible to create at the LHC. However, the long-term stability of a proton with a TeV-scale scalar LQ may be explained by some GUT formulations \cite{buchmuller1986constraints,dorsner2005unification,perez2007renormalizable}. In the lepton sector, recent studies have produced some intriguing findings that might point to an undiscovered BSM theory.  In addition, a joint finding from Brookhaven National Laboratory (BNL) and the Fermilab-based Muon $g-2$ team in 2021 revealed a $4.2\sigma$ difference between the measured and expected values of the anomalous magnetic moment of muon. In August 2023, the finding received an upgrade, increasing the significance to 5 $\sigma$ .

Color-triplet bosons that possess both lepton and baryon numbers are known as leptoquarks. Other quantum numbers are (fractional) electric charges having spins 0 (scalar LQ or "sLQ") or 1 (vector LQ or "vLQ"), depending upon the model under consideration. They do not exist in ordinary matter and, like the majority of other elementary particles, only survive for relatively brief periods. Also, they must carry a color charge, making it possible for them to interact with gluons. %If LQs exits, then its anti-particles with opposite charges also exit and decay into a conjugated state must exit as well. 
 In general, a leptoquark may interact with any combination of a quark and lepton with given electric charges (which yields up to $3\times3=9$, definite interactions of a single type of LQ).\\
  
 Various searches for LQs coupling to the first, second, and third generations have been published by the ATLAS and the CMS experiments \cite{iuppa2019searches,lo2021search,sirunyan2019search,aad2020search,aad2021search}. In our study, we investigate $3^{rd}$ generation scalar leptoquark pair production at LHC with a center of mass energy of $\sqrt{s}= 14$ TeV, focusing on their decay into top quarks and tau leptons in both hadronic and semileptonic channels.
 With distinct electric charges, each generation of leptoquarks is divided into up-type and down-type LQs.
 They are divided, for example, into down-type LQs ($LQ^{d}_{3}$), which decay into t$\tau$ or b$\nu$, and up-type LQs ($LQ^{u}_{3}$), which decay into b$\tau$ or t$\nu$ for the third generation.
 In our paper, we investigate $3^{rd}$ generation scalar leptoquark pair production involving semi-leptonic and hadronic decay modes.
 The LHC phenomenology of a scalar leptoquark with quantum numbers under the SM gauge group $S_{1}=(\bar{3},1,1/3)$ has been examined in this article. The leptoquark's existence also substantially improves the electroweak vacuum's stability \cite{bandyopadhyay2017vacuum}. Limits on the parameters of the various leptoquark model aspects are derived under the assumption of various branching fractions. A recent study at $13$ TeV data from CMS collaboration puts a bound on scalar $LQ$ of mass $\geq900~GeV$ in the search through t$\tau$ final states with $100\%$ branching fraction \cite{cms2018search}. The ATLAS Collaboration performed a cut-based analysis in a recent study \cite{atlas2020search} and derived the limit for the up-type third-generation scalar leptoquark, presuming that LQ decays to a top quark, as well as neutrino with a branching ratio of $100\%$.
 The leptoquark to t$\tau$ decays will be preferred over other decay modes by the third generation. Further, for the most current constraints on the vector leptoquark from collider searches, the references\cite{cms2020search,aad2023search} may be checked out. 
 %%%%%%%%%%%%%%%%%%%%%%%%%%%%%%%%%%%%%%%%%%%%%%%%%%%%%%
\begin{table}[!ht]
    \centering{}
    \begin{tabular}{|c|c|c|c|c|}
    \hline\hline
       $(SU (3), SU (2), U (1))$  & Spin & Symbol & Type & F \\
       \hline
        $(\Bar{3},3,1/3)$ & 0 & $S_{3}$ & LL($S^{L}_{1}$) & -2\\ \hline
        $(3, 2, 7/6)$ & 0 & $R_{2}$ & RL($S^{L}_{1/2}$),LR($S^{R}_{1/2}$) & 0 \\\hline
        $(3,2,1/6)$ & 0 & $\Tilde{R_{2}}$ & RL$(\Tilde{S}^{L}_{1/2}),\overline{LR}\Tilde{S}^{\overline{L}}_{1/2}$ & 0 \\\hline
        $(\Bar{3},1,4/3)$ & 0 & $\Tilde{S}_{1}$ & RR$(\Tilde{S}^{R}_{0})$ & -2 \\\hline
        $(\Bar{3},1,1/3)$ & 0 & $S_{1}$ & LL $(S^{L}_{0}),RR(S^{R}_{0}),\overline{RR}(S^{\overline{R}}_{0})$ & -2 \\ \hline
        $(\bar{3},1,-2/3)$ & 0 & $\Bar{S}_{1}$ & ${\overline{RR}}(\Bar{S}_{0}^{\overline{R}})$ & -2 \\ \hline
         \end{tabular}
    \caption{List of Scalar Leptoquarks}
    \label{scalar and vector leptoquarks of different coupling}
\end{table}
  %%%%%%%%%%%%%%%%%%%%%%%%%%%%%%%%%%%%%%%%%%%%%%%%%%%%%%
\subsection{LQ Model $S_{1}=(\bar{3},1,1/3)$}
 %%%%%%%%%%%%%%%%%%%%%%%%%%%%%%%%%%%%%%%%%%%%%%%%%%%%%%
The LQs that decay to a top quark and a charged lepton must have an electromagnetic charge of $\pm1/3$ according to electromagnetic charge conserving. S1 has been identified as a promising candidate for the decay mode, $LQ \rightarrow tl$, based on the classification of potential LQ states in Ref. \cite{dorvsner2016physics}.
The two renormalizable operators that are invariant under the SM gauge group ($G_{SM}$) for $S_1$ are as follows:
%\begin{eqnarray}
   % \mathcal{L} &\supset& + y^{LL}_{1~ij}\bar{Q}^{C~i,a}_{L} S_{1}\epsilon^{ab}L^{j,b}_{L} + y^{RR}_{1~ij}\bar{u}^{C~i}_{R} S_{1}e^{j}_{R} + y^{\overline{RR}}_{1~ij}\bar{d}^{C~i}_{R}S_{1}\nu^{j}_{R} \notag \\
  %  &+&z^{LL}_{1~ij}\bar{Q}^{C~i,a}_{L}S^{\ast}_{1}\epsilon^{ab}Q^{j,b}_{L} + z^{RR}_{1~ij}\bar{u}^{C~i}_{R}S^{\ast}_{1}d^{j}_{R} + h.c.
 %   \end{eqnarray}
  \begin{equation}
\mathcal{L} \supset y^{LL}_{ij} \, \bar{Q}^C_L{}^i S_1 i \tau^2 L_L{}^j + y^{RR}_{ij} \, \bar{u}^C_R{}^i S_1 e_R{}^j + \text{h.c.},
\end{equation}
where SM's left-handed quark and lepton doublets are denoted by $Q_L$ and $L_L$, accordingly.  Charge conjugation is indicated by the superscript C, which $\tau^{k}$ represents the Pauli matrices, where $k = {1,2,3}$.  The generation indices are represented by the notation $i, j = {1,2,3}$. This could possibly be stated clearly as:
\begin{equation}
\mathcal{L} \supset - (y^{LL}_i U)_{ij} \bar{d}^C_L{}_i S_1 \nu_L{}_j + (V^T y^{LL}_i)_{ij} \bar{u}^C_L{}_i S_1 e_L{}_j
+ y^{RR}_{ij} \bar{u}^C_R{}_i S_1 e_R{}_j + \text{h.c.}
\end{equation}
where U and V account for the Cabibbo-Kobayashi-Maskawa (CKM) quark mixing matrix and the Pontecorvo-Maki-Nakagawa-Sakata (PMNS) neutrino mixing matrix, accordingly.  We just designate the neutrino flavors with $\nu$ as they may not be discriminated at the LHC.  Similarly, the modest off-diagonal components with the CKM matrix have little impact on LHC phenomenology primarily and our study in particularly. For convenience, we thus use a diagonal CKM matrix. 
\begin{equation}
    \mathcal{L} \supset y^{LL}_{3j} \left( -\bar{b}^{C}_L \nu_L + \bar{\ell}^{C}_L \ell_L^{j} \right) S_1
    + y^{RR}_{3j} \bar{\ell}^{C}_R \ell_R^{j} S_1 + \text{h.c.},
\end{equation}
where $j ={1,2}$.

%Whereas $\tau^{\kappa}$ having $\kappa = 1,2,3$, are Pauli matrices $i,j = 1,2,3$ $(a,b = 1,2)$ are flavor $SU(2)$ indices, $\epsilon^{ab} = (\tau^2)^{ab}$, and the superscript $C$ stands for charge conjugation operator. An arbitrary complex $3\otimes3$, Yukawa coupling matrix has the entries $y^{LL}_{1~ij}$. In contrast, the $z_{1}^{LL}$  matrix is antisymmetric in flavor space, i.e., $z^{LL}_{1~ij} = -z^{LL}_{1~ij}$.
%In terms of quark-lepton and quark-quark pairings, the matrices $y$ and $z$ represent the strength of the LQ interaction.
%When the LQs only couple to left- or right-chiral quarks, respectively, they are referred to as left- and right-type LQs \cite{hirsch1996new}. In flavor space, $z^{LL}_{1}$ is a symmetric matrix (i.e $z^{LL}_{1~ij} = z^{LL}_{1~ji}$) \cite{davies1991tree}, while all other matrices are hypothetical and entirely arbitrary.\\
The $R_{D}^{*}$  anomalies may be well explained by LQs (often $S_{1}$) that pair with these fermions.
Also, it helps us to understand the anomalous magnetic moment of muon \cite{maes2022bold}.\\

%\begin{figure}[h]
 %   \centering
 %   \includegraphics[width=9cm,height=6cm]{l515.pdf}
  %  \caption{Production cross-section vs lambda for $3^{rd}$ Gen sLQ at different masses of LQ. }
   % \label{Crx}
%\end{figure}
%Fig.\ref{Crx} describes the relationship between $\lambda$ vs $\sigma$ for third-generation scalar leptoquark. As the coupling strength directly affects the probability of the leptoquark interacting with other particles, increasing it causes the cross-section to rise as well. A bigger cross-section in particle collisions indicates the probability of a certain interaction happening, and a higher coupling strength indicates that the leptoquark will interact with other particles more frequently.
 %%%%%%%%%%%%%%%%%%%%%%%%%%%%%%%%%%%%%%%%%%%%%%%%%%%%%%
\section{Benchmark points}
 %%%%%%%%%%%%%%%%%%%%%%%%%%%%%%%%%%%%%%%%%%%%%%%%%%%%%%
The searches for the ultimate states resulting from combinations of leptoquark decays to the third (t $\tau$) generations are the main topic of this article. We chose four benchmark points based on such decays, which are listed in Table \ref{B-Table}.
 %%%%%%%%%%%%%%%%%%%%%%%%%%%%%%%%%%%%%%%%%%%%%%%%%%%%%%

\begin{table}[h!]
\centering
\renewcommand{\arraystretch}{1.4}
\small % Reduces font size slightly to fit content
\begin{tabular}{|c|c|p{5cm}|c|c|}
\hline
\textbf{Benchmark Point} & \textbf{$\lambda_{33}$} & \textbf{Decay Channel (Path)} & \textbf{$M_{LQ}$ (GeV)} & \textbf{$\sqrt{s}$ (TeV)} \\
\hline
BP1 & 1.0 & $pp \rightarrow S_1 \bar{S}_1$, $S_1 \rightarrow t \tau^-$, $t \rightarrow b jj$, $\bar{S}_1 \rightarrow \bar{t} \tau^+$, $\bar{t} \rightarrow \bar{b} jj$ & 1350, 1500, 1650 & 14 \\
\hline
BP2 & 1.0 & $pp \rightarrow S_1 \bar{S}_1$, $S_1 \rightarrow t \tau^-$, $t \rightarrow bW$, $W \rightarrow \ell \nu_\ell$, $\bar{S}_1 \rightarrow \bar{t} \tau^+$, $\bar{t} \rightarrow \bar{b}W$, $W \rightarrow \ell \nu_\ell$ & 1350, 1500, 1650 & 14 \\
\hline
BP3 & 1.0 & $pp \rightarrow S_1 \bar{S}_1$, $S_1 \rightarrow t \tau^-$, $t \rightarrow b jj$, $\bar{S}_1 \rightarrow \bar{t} \tau^+$, $\bar{t} \rightarrow \bar{b} jj$ & 1500 & 14 \\
\hline
BP4 & 1.0 & $pp \rightarrow S_1 \bar{S}_1$, $S_1 \rightarrow t \tau^-$, $t \rightarrow bW$, $W \rightarrow \ell \nu_\ell$, $\bar{S}_1 \rightarrow \bar{t} \tau^+$, $\bar{t} \rightarrow \bar{b}W$, $W \rightarrow \ell \nu_\ell$ & 1500 & 14 \\
\hline
\end{tabular}
\caption{Benchmark points with detailed decay paths for fully hadronic and semi-leptonic channels in conventional cut-based method and machine-learning based algorithm at 14 TeV}
\label{B-Table}
\end{table}

%%%%%%%%%%%%%%%%%%%%%%%%%%%%%%%%%%%%%%%%%%%%%%%%%%%%%%%%%%%%%%%%
 \section{Production Cross-section}
 %%%%%%%%%%%%%%%%%%%%%%%%%%%%%%%%%%%%%%%%%%%%%%%%%%%%%%%%%%%%%%%
The parton-level leading-order (LO) cross sections are provided by \cite{hewett1988scalar} for a scalar leptoquark.
 \begin{eqnarray}
     \hat{\sigma}^{LO}_{gg}&=&\frac{\alpha^{2}_{s}\pi}{96\hat{s}} \times [\beta(41-31\beta^{2}) + (18\beta^{2}-\beta^{4}-17)log\frac{1+\beta}{1-\beta}] \nonumber\\
 \hat{\sigma}^{LO}_{q\Bar{q}} &=& \frac{2\alpha^{2}_{s}\pi}{27\hat{s}}\beta^{3}\nonumber\\
 \hat{\sigma}^{LO}_{eq} &=& \frac{\pi\lambda^{2}}{4} \delta(\hat{s}-M^{2}_{LQ})\nonumber
 \end{eqnarray}
 Here, the parton subprocess's invariant energy is represented by $\Hat{s}$, and $\beta = \sqrt{1-4M^{2}_{LQ}}/{\hat{s}}$ .
 %%%%%%%%%%%%%%%%%%%%%%
\begin{figure}[h]
 	\centering
   \subfigure[]{\label{fig:a}\includegraphics[width=80mm]{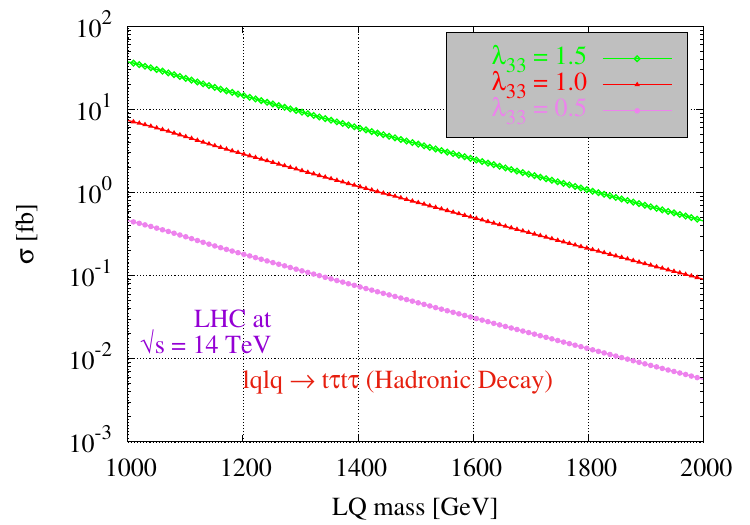}}
    \subfigure[]{\label{fig:b}\includegraphics[width=80mm]{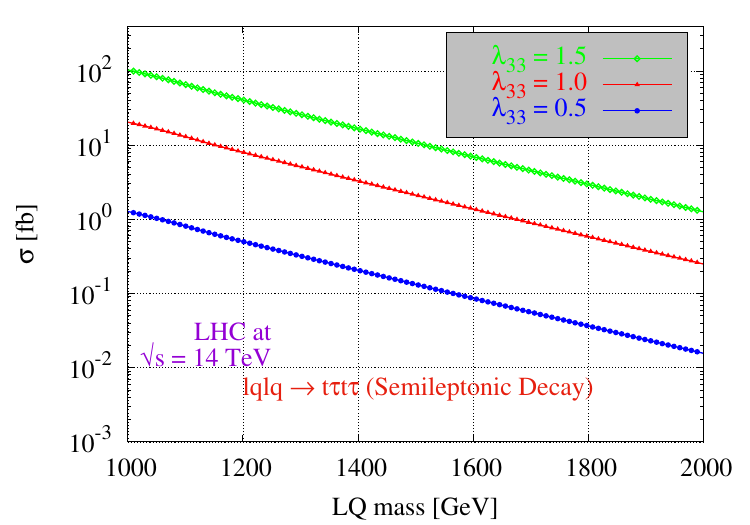}}
	  \caption{Cross-section as a function of mass for third generation sLQ (a) correspond to the hadronic decay channel and (b) illustrates semileptonic channel at different values of lambda's.}
	 \label{mass-crx}
\end{figure}

 %%%%%%%%%%%%%%%%%%%%%%
 Fig.\ref{mass-crx} shows the variation of cross section with  sLQ mass at $\sqrt{s}= 14~ TeV$ for both hadronic and semileptonic channels. We found that, semileptonic decay mode yields a bit higher production cross section than hadronic one. Because the presence of isolated lepton in the semileptonic mode provides clean trigger while fully hadronic final states encounter large QCD backgrounds might decrease the net yields. 
 The leptoquark pair-production cross-section is known to be essentially independent of the Yukawa couplings except for extremely high values.\\
The production cross-section of a pair-produced leptoquark in femto barn (fb) with the leptoquark mass fluctuation at the 14 TeV, LHC is displayed in Figure \ref{mass-crx}. The phase space that may be used to produce leptoquarks gets smaller as their mass rises. Because of this, there is a lower chance of creating the leptoquark in high-energy collisions. The likelihood of producing the leptoquark in collisions is decreased by this lower connection. The production of the leptoquark becomes difficult at the greater masses, because of the more severe energy and momentum limits in the collision.\\
%%%%%%%%%%%%%%%%%%%%%%%%%%%%%%%%%%%%%%%%%%%%%%%%%%%%%%%%%%%%%%%
\begin{figure}[!ht]
    \centering
    \includegraphics[width=9cm,height=6cm]{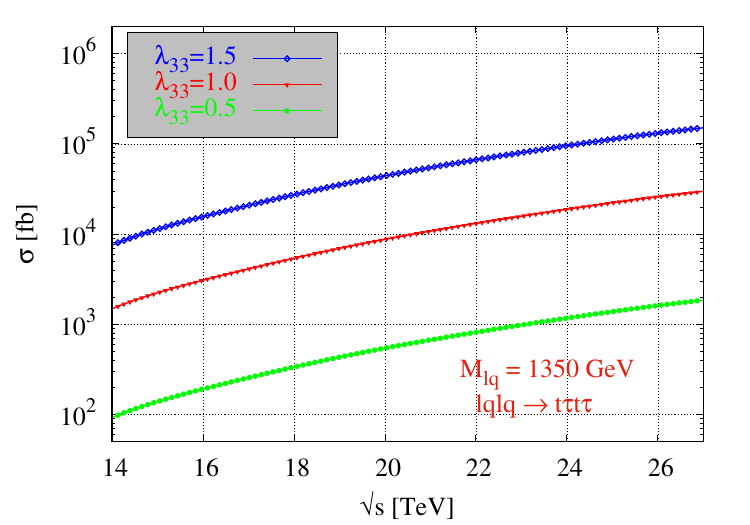}
    \caption{Energy vs cross-section for third generation pair-production sLQ at different values of lambda's.}
    \label{E-crx}
\end{figure}

Accordingly, it is more likely for the particles to have sufficient energy to generate leptoquarks when the level of energy rises.\\
%%%%%%%%%%%%%%%%%%%%%%%%%%%%%%%%%%%%%%%%%%%%%%%%%%%%%%%%%%%%%%%
Figure \ref{E-crx} shows the relationship between energy vs cross-section at different couplings. In particle collisions involving leptoquarks, the cross-section grows as the center of mass energy increases. Via exchanging kinetic energy, heavier particles may be created at a higher center of mass energies. This leads to a greater cross-section by increasing the number of alternative end states. More phase space is available for particle generation at the higher energies. This increases the frequency of interactions and, consequently, the cross-section since there are more possibilities for the particles involved in the collision to organize themselves. Specific particle resonant generation is more likely at the higher energies. The cross-section will grow as the energy gets closer to the resonance energy if the leptoquark manufacturing process is connected to any resonant states.

Total decay widths of leptoquarks result in a pair of leptons and quarks \cite{belyaev2005leptoquark}
\begin{equation}
    \Gamma_{j=0} = \sum_{i}\dfrac{\lambda^{2}_{i}}{16\pi} M_{LQ}
\end{equation}
where $j$ here indicates the spin of the leptoquark, and the sum is extended to all possible decay modes of the leptoquark.  %Leptoquarks can decay either to charged leptons and quarks or to neutrinos and quarks, depending on the type of leptoquark.
%The above equation shows the total decay width for scalar leptoquarks. We have to calculate partial decay width, for this, we have to divide partial decay width by total decay width \ref{lamda}.  
%%%%%%%%%%%%%%%%%%%%%%%%%%%%%%%%%%%%%%%%%
\begin{figure}[h]
    \centering
    \includegraphics[width=9cm,height=6cm]{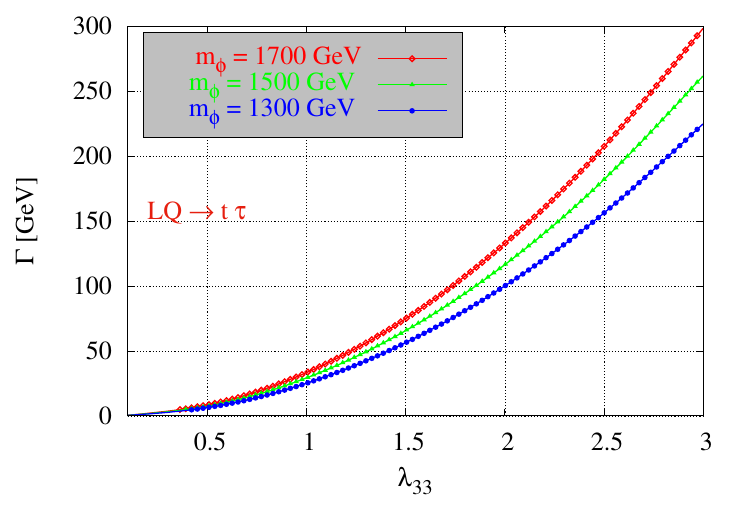}
    \caption{Total decay width vs lambda for third generation sLQ at different masses. }
    \label{lamda}
\end{figure}
%%%%%%%%%%%%%%%%%%%%%%%%%%%%%%%%%%%%%%%%%
All particles, including leptoquarks, tend to decay wider overall as their mass increases. Higher mass particles may be permitted to exist in a wider range of energy states, which might result in a wider variety of decay modes and a wider overall decay breadth. This indicates that there are often more decay channels available for the heavier particles. A wider decay width results from an increase in the number of available decay channels caused by an increase in the coupling between a leptoquark and other particles. This increases the chance of decay and, thus, the decay width of the leptoquark, since it can decay into a greater range of particles.

Integrated luminosity is calculated as;  
\begin{equation}
    N = \sigma \int~Ldt = \sigma L_{int},
\end{equation} 
%%%%%%%%%%%%%%%%%%%%%%%%%%%%%%%%%%%%%%%%%
\begin{figure}[!ht]
    \centering
    \includegraphics[width=9cm,height=6cm]{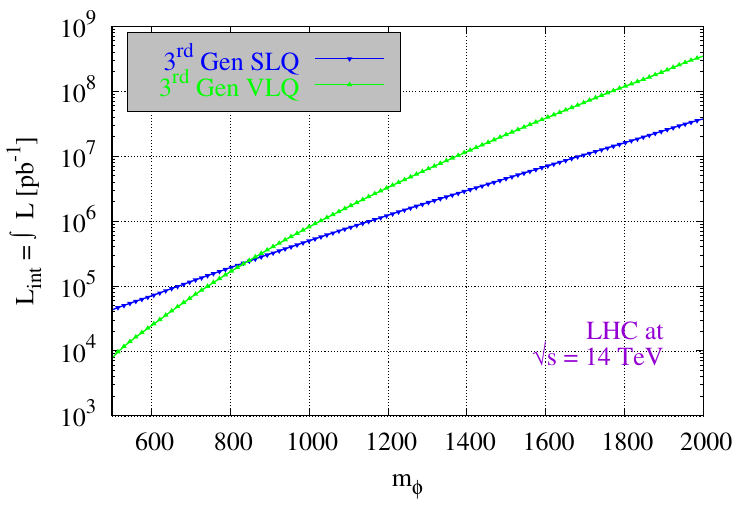}
    \caption{Required integrated luminosity for scalar and vector leptoquarks. }
    \label{lumi}
\end{figure}
%%%%%%%%%%%%%%%%%%%%%%%%%%%%%%%%%%%%%%%%%
In our investigation, we found that there is a relationship between the mass of leptoquarks and the integrated luminosity. In particular, the integrated luminosity, which is determined by dividing the number of observed events by the cross-section, tends to grow when we consider increasing leptoquark masses.
The correlation between the mass of the particles and the cross-section can be used to explain this tendency. With increasing mass, the production cross-section for leptoquarks gradually declines. Lower production of leptoquarks in collisions with larger masses results in fewer events being reported. However, this is counteracted by dividing the total number of events by the smaller cross-section in the integrated luminosity estimate.
As a result, at higher masses, when the cross-section is smaller, the integrated luminosity value increases to account for the additional experimental work required to witness a substantial number of events. This emphasizes how harder it is to find greater mass leptoquarks and how crucial it is for the investigations to get higher integrated luminosities to maintain sensitivity throughout a wide mass range.
%%%%%%%%%%%%%%%%%%%%%%%%%%%%%%%%%%%%%%%%%%%%%%%%
\section{Kinematical Analysis}
%%%%%%%%%%%%%%%%%%%%%%%%%%%%%%%%%%%%%%%%%%%%%%%%
In this section, we present a kinematical analysis of leptoquarks production decay modes via calculations on invariant mass spectra, transverse momenta, pseudorapidity, and signal significance. Using MadGraph5 \cite{alwall2014automated}, the simulated sample for leptoquarks was produced at a center-of-mass energy of 14 TeV and at leading order using the NNPDF23 LO PDF set (nn23lo1) \cite{ball2013parton}. Similarly, the events for both the signal and the background are generated using MadGraph v3.4.5, and these events are then written to an LHE file and compiled using PYTHIA v3.4.5. Additionally, PYTHIA determines their relative efficiency. The results are subsequently analyzed using ROOT v6.26.02. The histograms and graphs are generated as well using ROOT analysis toolikt \cite{brun1997root}.\\

Different scattering processes are employed as signals in the present research. The LHC p-p collisions at $\sqrt{s}$=14 TeV produce all of these scattering modes. However, protons also include the charged leptons because of quantum fluctuations, making it feasible to analyze lepton-driven events in the LHC. The collision of a lepton ($\ell$) from one proton and a quark (q) from the other proton results in the resonant generation of an exotic leptoquark (LQ) state, which is the most basic example of this type of mechanism. To enhance the simulation findings on certain kinematical cuts in this work, 80000 events are created and integrated for the signal processing.\\

Integrated luminosities 300 $fb^{-1}$,  1000 $fb^{-1}$ and 3000 $fb^{-1}$ are used to calculate the signal prominence. Using the momenta and energies of its decay products, the invariant mass of a leptoquark is a basic feature that measures the overall mass energy of the leptoquark system. Invariant mass is an essential tool for understanding the characteristics and interactions of leptoquarks in particle physics studies as it is constantly independent of the observer's frame of reference. In the signal processes, the following final states are possible for third-generation scalar leptoquark: 
\begin{eqnarray}
        pp&\rightarrow& S_1 \Bar{S_{1}},~S_{1}\rightarrow t\tau,~\Bar{S_{1}}\rightarrow \Bar{t}\Bar{\tau}, t\rightarrow b jj, \Bar{t} \rightarrow \Bar{b} jj
        \label{3}\\
        pp&\rightarrow& S_{1}\Bar{S_{1}},~S_{1}\rightarrow t\tau,~\Bar{S_{1}}\rightarrow \Bar{t}\Bar{\tau}, t\rightarrow b W, \bar{t} \rightarrow \bar{b} W, W \rightarrow l~  \nu l
        \label{4}
        \end{eqnarray}
While evaluating these signal circumstances, we took into account the SM background processes listed in Table \ref{SM-B} with comparable end-state topologies. 
The signal significance is calculated by using the cuts as follows:
\begin{table}[h!]
\centering
\begin{tabular}{|c|c|c|}
\hline
\textbf{Selection Variable} & \textbf{Semileptonic Channel} & \textbf{Fully Hadronic Channel} \\
\hline
$E_T^{\text{miss}}$ & $>$ 500 GeV & $>$ 250 GeV \\
\hline
Number of jets ($N_{\text{jets}}$) & $\geq$ 4 & $\geq$ 6 \\
\hline
Total Hadronic Transverse Energy ($H_T$) & $>$ 1200 GeV & $>$ 1500 GeV \\
\hline
\multicolumn{3}{|c|}{\textbf{Common Cuts for Both Channels}} \\
\hline
$p_T$ (jet) & \multicolumn{2}{c|}{$>$ 700 GeV} \\
\hline
$p_T$ ($\tau$ lepton) & \multicolumn{2}{c|}{$>$ 120 GeV} \\
\hline
$p_T$ (electron/muon) & \multicolumn{2}{c|}{$>$ 50 GeV} \\
\hline
$|\eta|$ (lepton, jet) & \multicolumn{2}{c|}{$<$ 2.5} \\
\hline
Number of $\tau$ leptons & \multicolumn{2}{c|}{$\geq$ 1} \\
\hline
$\Delta R(j,j)$ & \multicolumn{2}{c|}{$>$ 0.4} \\
\hline
$\Delta R(\ell,j)$ & \multicolumn{2}{c|}{$>$ 0.4} \\
\hline
\end{tabular}
\caption{Cut-based selection criteria applied in \texttt{MadAnalysis5} for semileptonic and fully hadronic channels.}
\label{tab:cutflow}
\end{table}

\begin{figure}[h]
 	\centering
   \subfigure[]{\label{fig:a}\includegraphics[width=80mm]{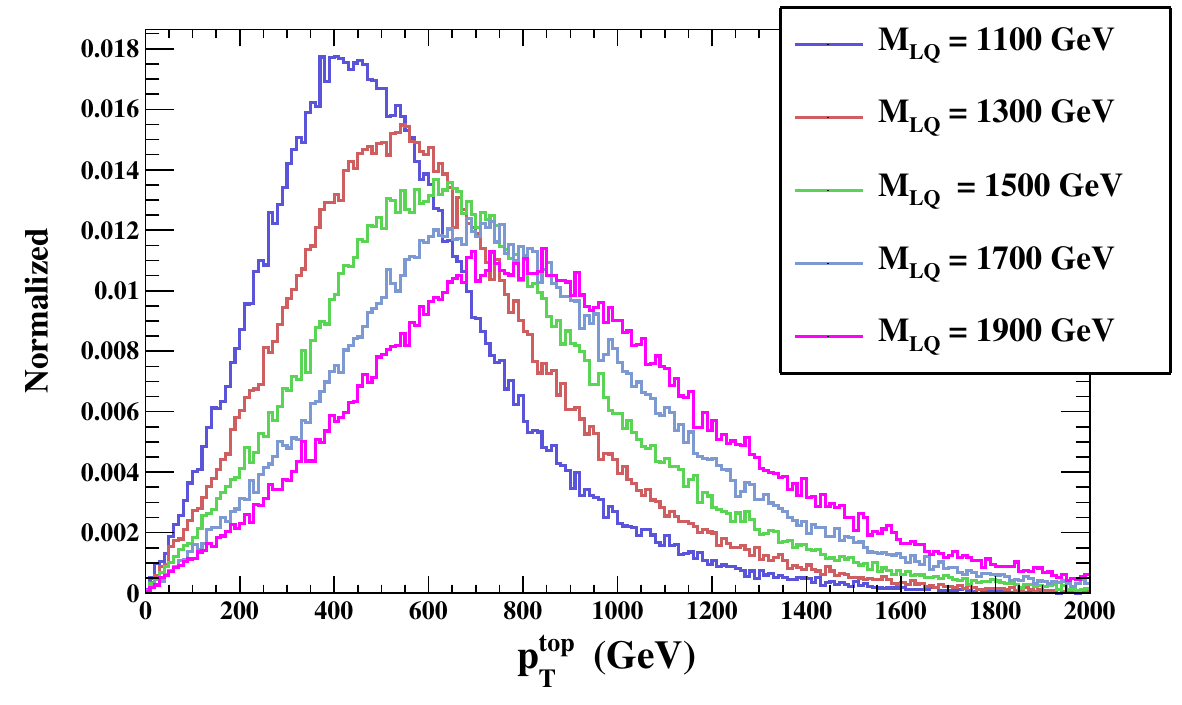}}
    \subfigure[]{\label{fig:b}\includegraphics[width=80mm]{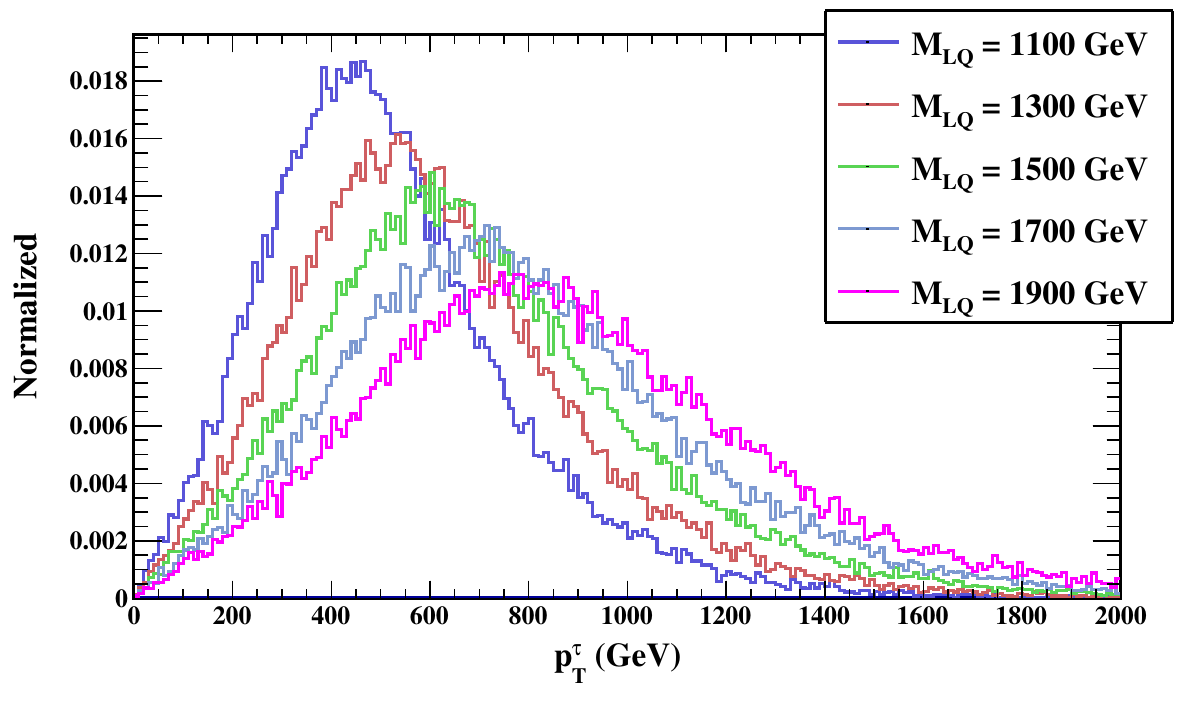}}
	\caption{Transverse momenta distributions of (a) top quarks and (b) charged leptons that result directly from scalar LQ decay simulations.} 
	\label{toptau}
\end{figure}

\begin{figure}[h]
    \centering
    \includegraphics[width=9cm,height=6cm]{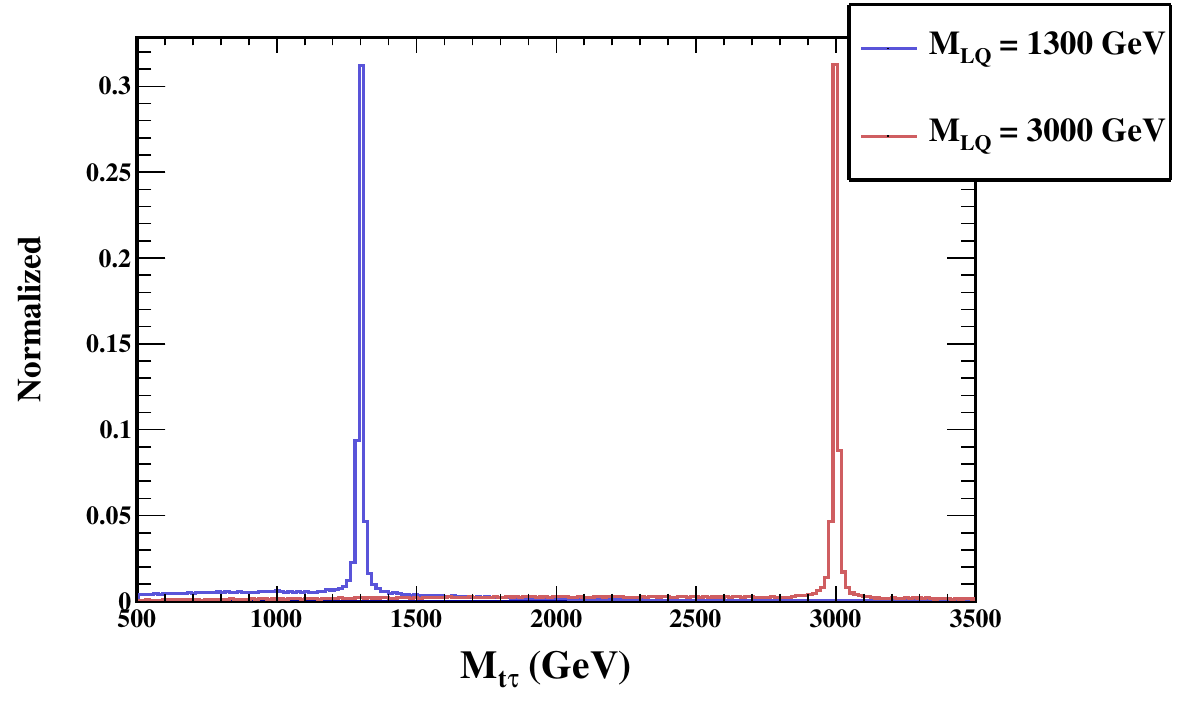}
    \caption{Invariant mass of leptoquark from top-tau pair.}
    \label{INV MASS LHE}
\end{figure}

The initial observations in high-energy physics experiments are transverse momentum distributions of particles of the final state. Statistical techniques have been frequently used in recent years to explain such multi-particle systems. 
%A
%particle's transverse momentum can be described as:
%\begin{equation}
	%P_{T} = \sqrt{P_{x^{2}} + P_{y^{2}}}
%\end{equation}
%where the momentum variables in the transverse momentum axis are $P_{x}$ and $P_{y}$. In terms of transverse energy, it is defined as:
%\begin{equation}
	%E_{T} = \sqrt{M^{2} + P^{2}_{T}}
%\end{equation}
Large transverse momenta of the particles created in the LQ decays are often assumed to characterize the signal processes. However, as Figure \ref{toptau} demonstrates exemplarily for the transverse momenta of top quarks and charged leptons arising from the decay $LQ\rightarrow t\tau$, the kinematic behavior of the decay products depends extensively on the LQ mass.
Pseudorapidity is characterized in  terms of coordinate system \cite{austrup2022search} as:
\begin{equation}
    \eta =  -\ln (\tan\dfrac{\theta}{2})  
\end{equation}
%\begin{equation}
	%y = \dfrac{1}{2} \ln(\dfrac{E + p_{z}}{E - p_{z}}) =  \dfrac{1}{2} \ln\dfrac{1 + \cos{\theta}}{1 -\cos{\theta}} = \ln %\cot\dfrac{\theta}{2} = \eta
%\end{equation}
which in the relativistic limit corresponds to the rapidity, y, ascertained from the momentum of a particle along the beam axis, $p_{z}$, and its energy, E.
\begin{equation}
    y = \dfrac{1}{2} \ln(\dfrac{E + p_{z}}{E - p_{z}})
\end{equation}
Where $\eta$ is the pseudorapidity. 
%%%%%%%%%%%%%%%%%%%%%%%%%%%%%%%%%%%%%%%%%%%%%%%%
\begin{figure}[h]
 	\centering
   \subfigure[]{\label{fig:a}\includegraphics[width=80mm]{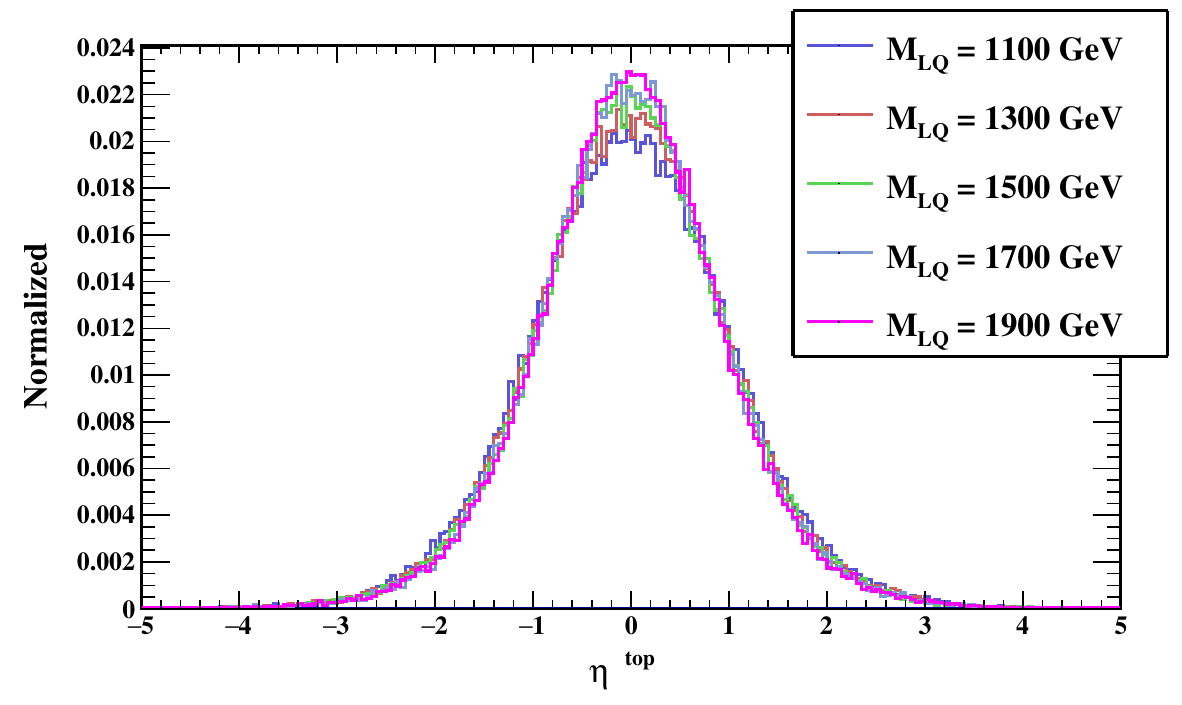}}
    \subfigure[]{\label{fig:b}\includegraphics[width=80mm]{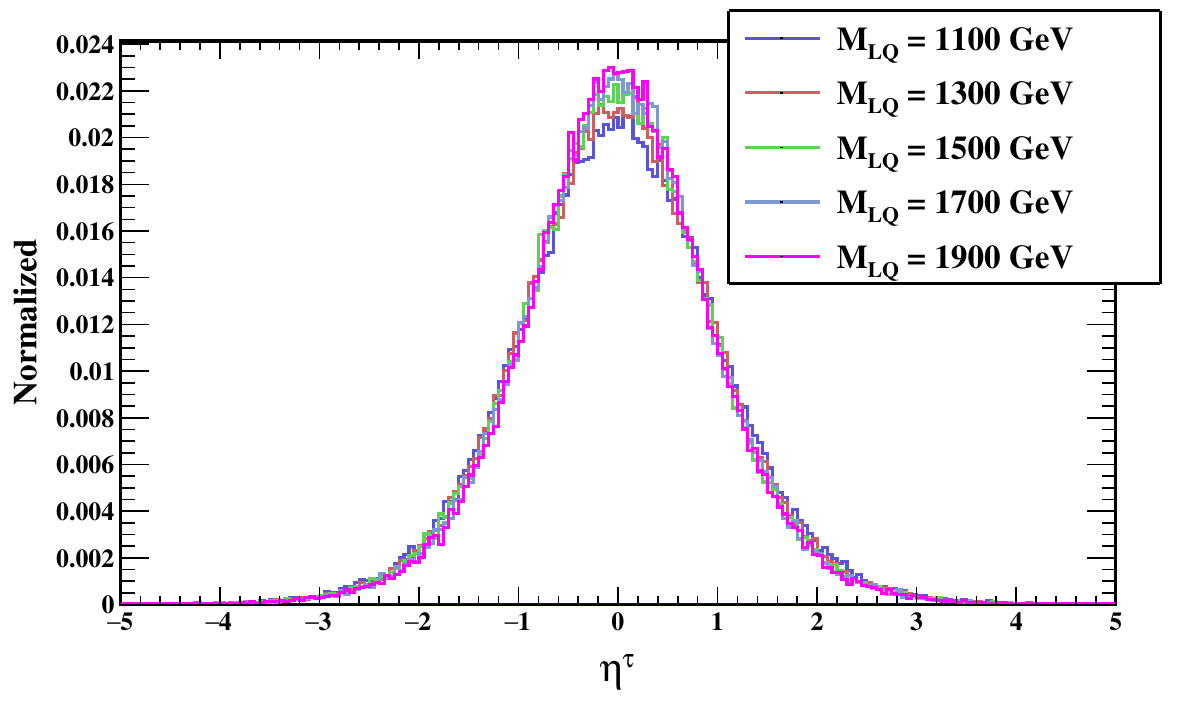}}
	\caption{Pseudorapidity of top quark on left, while for $\tau$ on right respectively.} 
	\label{INVH}
\end{figure}
%%%%%%%%%%%%%%%%%%%%%%%%%%%%%%%%%%%%%%%%%%%%%%%%
By definition, a leptoquark has the unique ability to decay after creation into a quark and a lepton. At the LHC, the quark eventually transforms into a hadronic jet, and we can rebuild the leptoquark resonance using the invariant mass of the jet-lepton pair. The invariant mass distribution of one top and one tau can be found in Figure \ref{INV MASS LHE}, where a clear peak can be seen at 1300 GeV and 3000 GeV. In Figure \ref{INVH}, we present the invariant mass distributions from the various quark flavors at 1300 GeV, simulated at the $\sqrt{s}$ = 14 TeV for LHC.
%%%%%%%%%%%%%%%%%%%%%%%%%%%%%%%%%%%%%%%%%%%%%%%%
%%%%%%%%%%%%%%%%%%%%%%%%%%%%%%%%%%%%%%%%%%%%%%%%
\begin{figure}[h]
 	\centering
   \subfigure[]{\label{fig:a}\includegraphics[width=80mm]{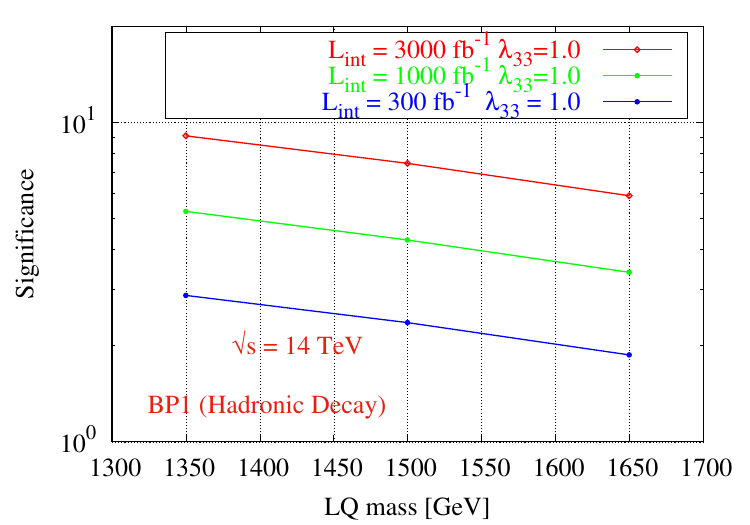}}
    \subfigure[]{\label{fig:b}\includegraphics[width=80mm]{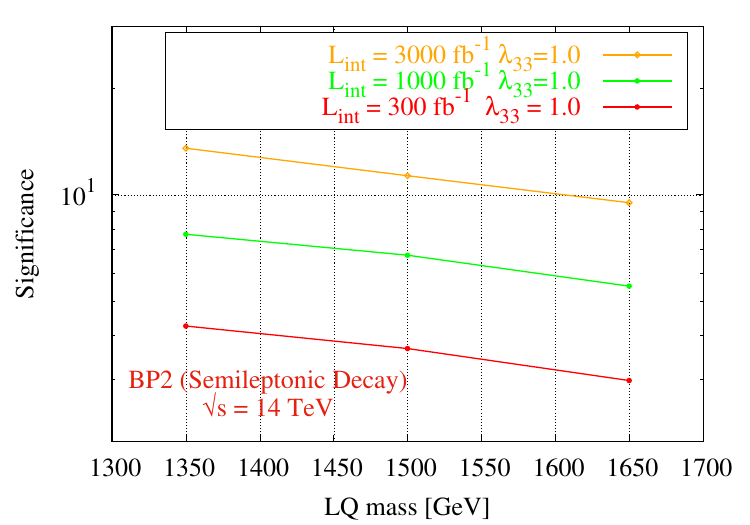}}
	\caption{Mass vs Significance for BP1 and BP2.} 
	\label{sig}
\end{figure} 
%%%%%%%%%%%%%%%%%%%%%%%%%%%%%%%%%%%%%%%%%%%%%%%%
\begin{figure}[h]
 	\centering
   \subfigure[]{\label{fig:a}\includegraphics[width=80mm]{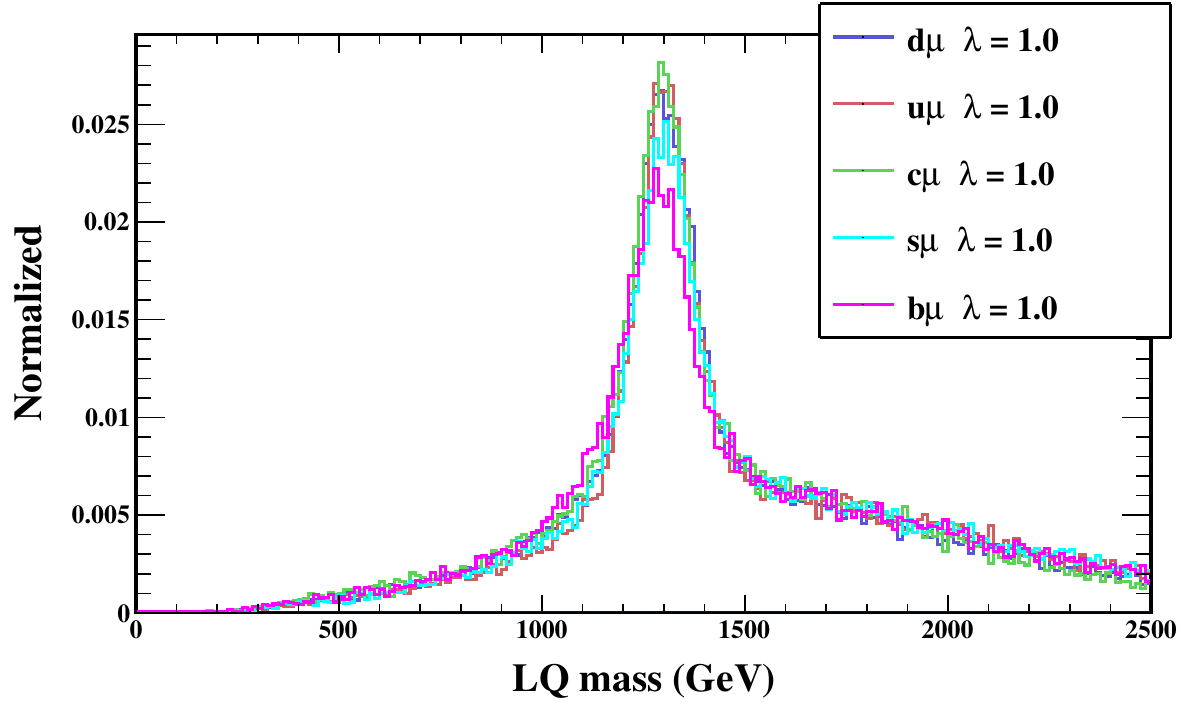}}
    \subfigure[]{\label{fig:b}\includegraphics[width=80mm]{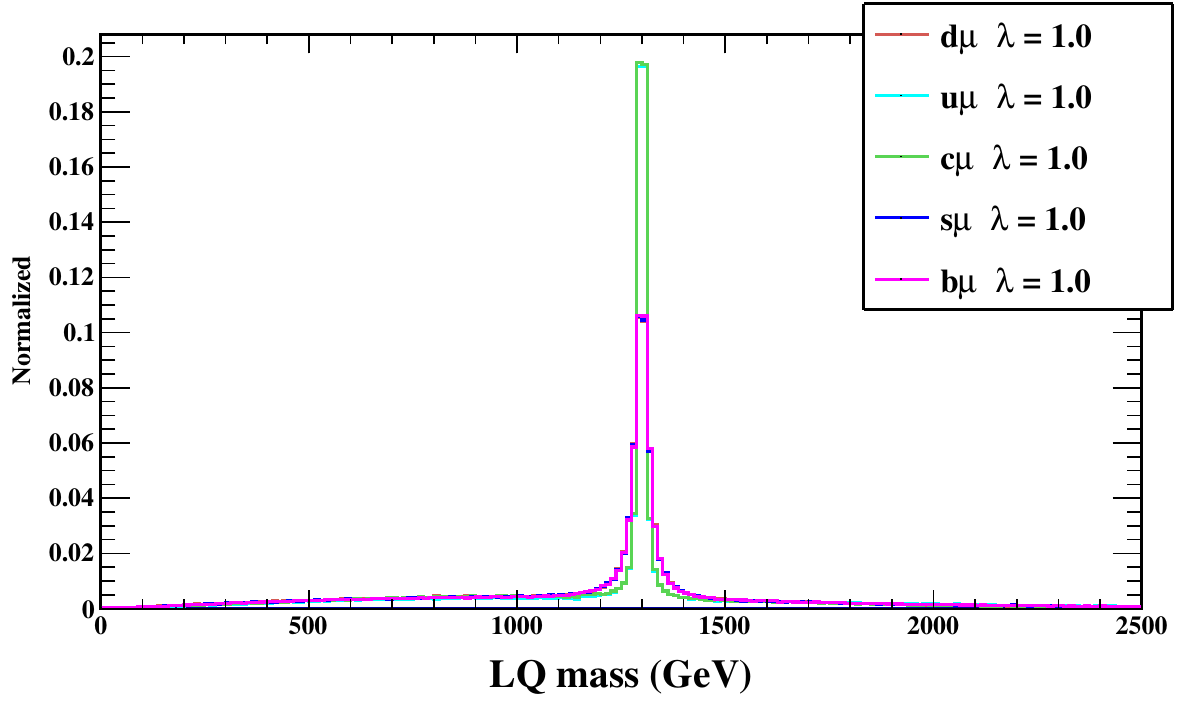}}
	\caption{Invariant mass of leptoquark for different quark flavour at $\sqrt{s}$ = 14 TeV.} 
	\label{sig}
\end{figure} 
%%%%%%%%%%%%%%%%%%%%%%%%%%%%%%%%%%%%%%%%%%%%%%%%
Fig. \ref{sig} highlights the mass (LQ) vs significance for various integrated luminosities. It should be mentioned that, with increasing mass of leptoquark, the significance decreases  with higher mass, the cross-section decreases. The completely hadronic decay mode has a bit smaller significance than the semileptonic channel due to its huge QCD backgrounds, even if greater integrated luminosities increase the statistical significance. Because isolated leptons are present, the semileptonic decay mode has cleaner event selection, which leads to a much larger significance for all mass values.
Greater integrated luminosity provides more information to precisely estimate and minimize background contributions. As a result, better background removal methods are made possible, producing a more efficient signal with greater significance. Also, the experiment's sensitivity to rare phenomena, including the creation and decay of the third-generation scalar leptoquark, is increased by the integrated luminosity. This increases its relevance by allowing signal events that were previously hidden by the background noise to be detected.
%%%%%%%%%%%%%%%%%%%%%%%%%%%%%%%%%%%%%%%%%%%%%%%%%%
\section{Multivariate Analysis}
%%%%%%%%%%%%%%%%%%%%%%%%%%%%%%%%%%%%%%%%%%%%%%%%%%
The experimentalists had to optimize their information extraction processes by using the most effective multivariate analysis techniques because of the inherent complexity of the data they collected and their desire to properly use the knowledge they provided on subnuclear phenomena. The LHC experiments were able to accomplish that goal by extracting more information from their datasets and enhancing the calibre of their scientific output by the concurrent development of contemporary machine learning (ML) techniques and the growing extent of their application to scientific research. In this work for the implementation of machine learning techniques, the ``Toolkit for Multivariate Analysis (TMVA)", a ROOT-integrated framework, is used that allows for the processing and calculation of many multivariate classification algorithms simultaneously \cite{brun1997root}. 

%%%%%%%%%%%%%%%%%%%%%%%%%%%%%%%%%%%%%%%%%%%%%%
 \subsection{Input Variables and Methods Used for Analysis}
 %%%%%%%%%%%%%%%%%%%%%%%%%%%%%%%%%%%%%%%%%%%%%%
We used seven classifiers in our work named as:
  Boosted Decision Tree (BDT), Likelihood, Multilayer Perceptron (MLP), LikelihoodD,
  BDT Discreminent and Boosted Fisher.\\ 
To choose the best classifier with the implementation of cut-based analysis, we have used the following input variables for all classifiers:\\
Jet transverse momentum (Jet $P_{T}$), Jet Pseudorapididity ($\eta_{jet}$), Number of jets ($N_{jet}$), Jet radius ($\Delta{R}_{jet}$), Scalar sum of all $P_{T}$ (Scalar$H_{T}$)
and Missing energy ( $\cancel{E_{T}}$ ).
The description of classifiers is discussed in detail in below sections.
 %%%%%%%%%%%%%%%%%%%%%%
\subsection{BDTs (Boosted Decision Trees)}
 %%%%%%%%%%%%%%%%%%%%%%
A classification tool with a tree-like topology is called a decision tree. Repetitive left/right or (yes/no) selections are made per single test event, one variable at a time, until the event reaches a node known as the \enquote{leaf node}, which categorizes it as either background or signal. The phase space is separated into several distinct areas by the collection of leaf nodes, which are categorized as either signals or background types. \\
A modification of a single decision tree is represented by boosted decision trees. Individual decision tree classifications are integrated to create an ensemble of decision trees, and the classifier is determined by a majority (weighted)  vote of each of the decision tree classifications.
 %%%%%%%%%%%%%%%%%%%%%%
\subsection{Likelihood Ratio}
 %%%%%%%%%%%%%%%%%%%%%%
Constructing an algorithm of one-dimensional probability density functions (PDFs) from the training data that replicates the data parameters for signal and background is the foundation for the maximum likelihood categorization method. By multiplying the signal probability densities of each input variable and normalizing the result by the total of the signal and background likelihoods, one may determine the likelihood that a particular occurrence corresponds to a given signal format.
The probability ratio $y_{L} (j)$ for occurrence j is determined by
\begin{equation}
    y_{L}(j) = \dfrac{L_{S}(j)}{L_{S}(j) + L_{B}(j)}
    \label{likelihood ratio}
\end{equation}
In addition, the following equation could be used to determine a candidate's likelihood of being signal or background:
\begin{equation}
    L_{S/B}(j) =\prod_{i=1}^{n_{var}}P_{S/B,i} (x_{i} (j))
    \label{likelihood S/B}
\end{equation}
while the probability density function for the ith input variable, $x_{i}$, is represented by $P_{S/B, i}$. For every i, the probability density function gets normalized at one:

\begin{equation}
     \int_{\infty}^{-\infty}P_{S/B,i} (x_{i})dx_{i} = 1
     \label{PDF normalization}
\end{equation}
 %%%%%%%%%%%%%%%%%%%%%%%%%%%%%%%%%%%%%%%%%%%%%%%%%%%%%%%%
\subsection{Multilayer Perceptron}
\begin{figure}[h]
    \centering
    \includegraphics[width=105mm]{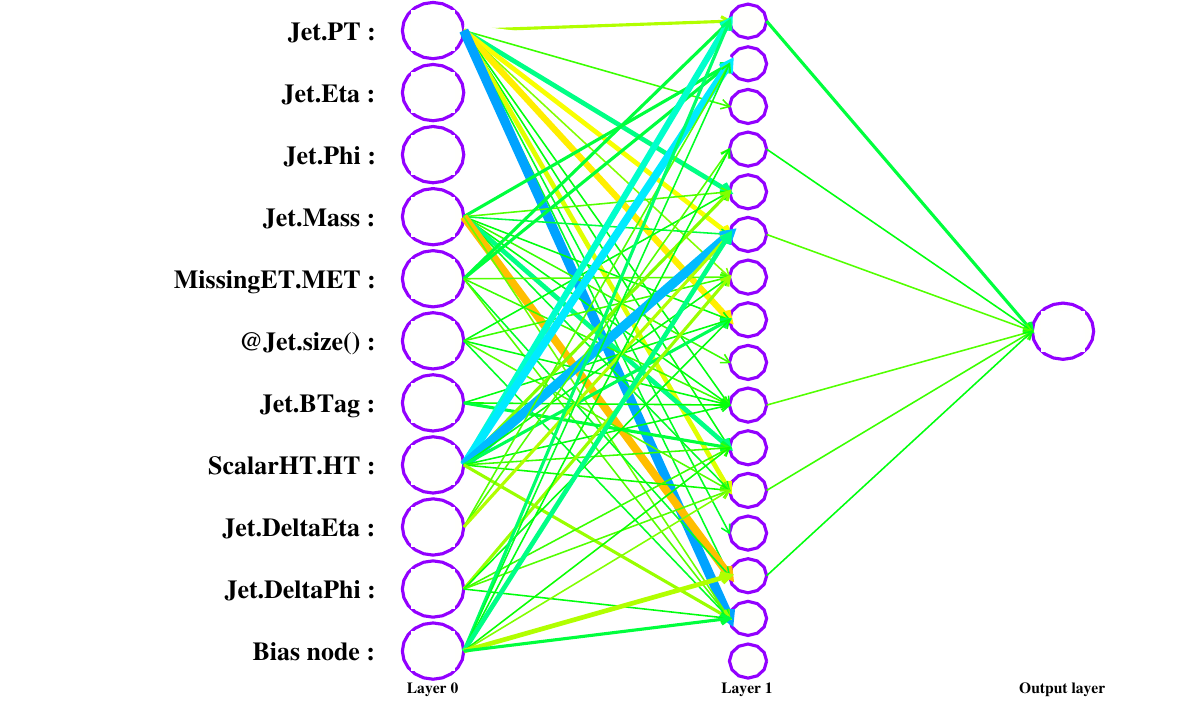}
    \caption{Layout of Neural network}
    \label{mlp}
\end{figure}

 %%%%%%%%%%%%%%%%%%%%%%%%%%%%%%%%%%%%%%%%%%%%%%%%%%%%%%%%
An architecture known as a Multilayer Perceptron, or neural network, is made up of many hidden layers of neurons, with each layer's output serving as the input for the subsequent layer's neurons, as seen in Figure \ref{mlp}. Recurrent neural networks are examples of additional linkages that allow a neuron's output to be utilized as another neuron's input within a single layer. A binary classification tool or a multi-class classifier can be used to separate the signal from the noise \cite{choi2020introduction}.
 %%%%%%%%%%%%%%%%%%%%%%%%%%%%%%%%%%%%%%%%%%%%%%%%%%%%%%%%
 %%%%%%%%%%%%%%%%%%%%%%%%%%%%%%%%%%%%%%%%%%%%%%%%%%%%%%%%
\subsection{Fisher discriminants (Linear Discriminant Analysis)} 
 %%%%%%%%%%%%%%%%%%%%%%%%%%%%%%%%%%%%%%%%%%%%%%%%%%%%%%%%
By separating the mean values of the signal and background distributions, event selection is carried out in a transformed variable space with zero linear correlations using the Fisher discriminants approach. The linear discriminant analysis identifies an axis in the (correlated) hyperspace of the input variables, such that events of the same class are confined to a close area, and the output classes (background and signal) are pushed as far apart as possible when projected onto this axis. The measure used to define "far apart" and "close vicinity" is the covariance matrix of the discriminating variable space, which reflects the classifier's linearity quality.
The following traits are used to categorise the occurrences into signal and background classes: \\
For each input variable $k = 1,...,n_{var}$, the overall sample means $\bar{x}_{k}$, the class-specific sample means $\bar{x}_{S(B),k}$, and the sample's total covariance matrix C. One way to break down the covariance matrix is to add the \textit{within-class (W)} and \textit{between-class matrix (B)}.  Both of them explain how occurrences are distributed in relation to the means of their individual classes (within-class matrix) and the sample means as a whole (between-class matrix). After that, the \textit{Fisher coefficients}, $F_{k}$, are provided by:
\begin{equation}
    F_{k}=\dfrac{\sqrt{N_{S}N_{B}}}{N_{S}+N_{B}}\sum_{\ell=1}^{n_{var}}W^{-1}_{k\ell}(\bar{x}_{S,\ell}-\bar{x}_{B,\ell})
\end{equation}
where $N_{S(B)}$ is the training sample's signal (background) event count. For event $i$, the Fisher discriminant $y_{F_{i}}(i)$ is provided by:
\begin{equation}
    y_{F_{i}}(i)=F_{o}+\sum_{\ell=1}^{n_{var}} F_{k}x_{k}(i)
\end{equation}
The sample mean $y_{F_{i}}$ of all $N_{S} + N_{B}$ occurrences is centred at zero by the offset $F_{o}$. Instead of using the \textit{within-class matrix}, the Mahalanobis variant determines the Fisher coefficients as follows:
\begin{equation}
     F_{k}=\dfrac{\sqrt{N_{S}N_{B}}}{N_{S}+N_{B}}\sum_{\ell=1}^{n_{var}}C^{-1}_{k\ell}(\bar{x}_{S,\ell}-\bar{x}_{B,\ell})
\end{equation}
Here $C_{k\ell}=W_{k\ell}+B_{k\ell}$.
%%%%%%%%%%%%%%%%%%%%%%%%%%%%%%%%%%%%%%%%%%%%%%%%%%%%%%%%
\subsection{Signal Efficiency versus Background Rejection}
\begin{figure}[!ht]
 	\centering
   \subfigure[]{\label{fig:a}\includegraphics[width=81mm,height=5cm]{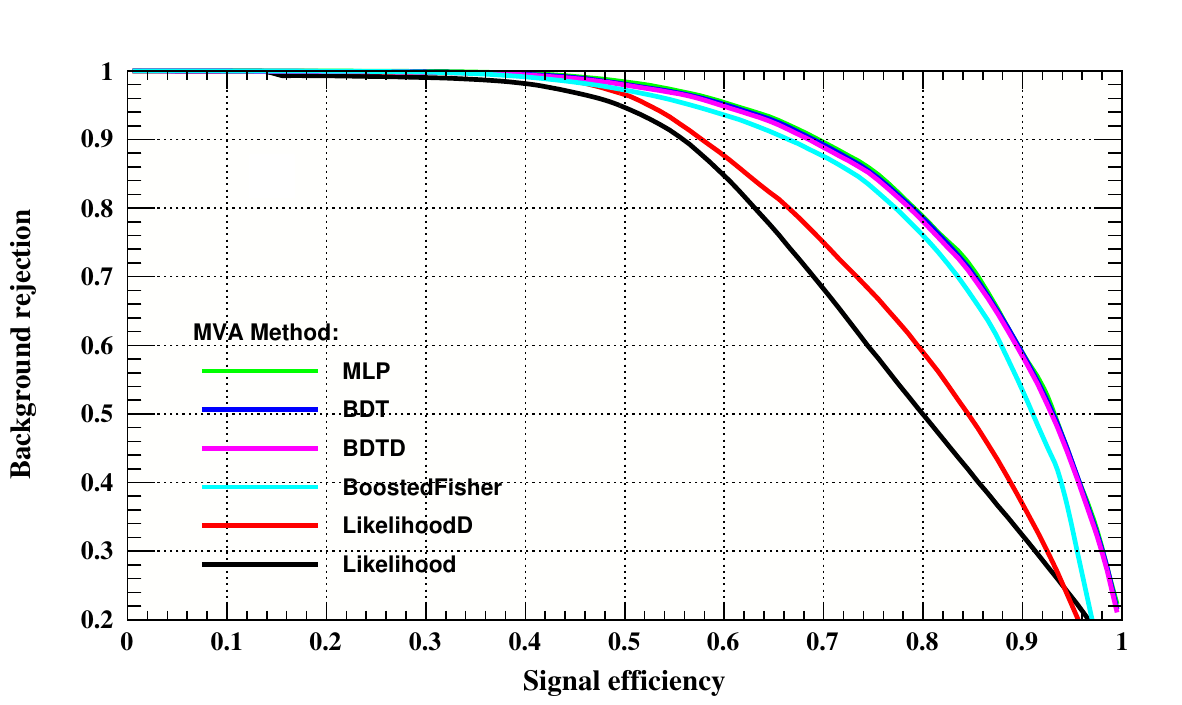}}
    \subfigure[]{\label{fig:b}\includegraphics[width=81mm,height=5cm]{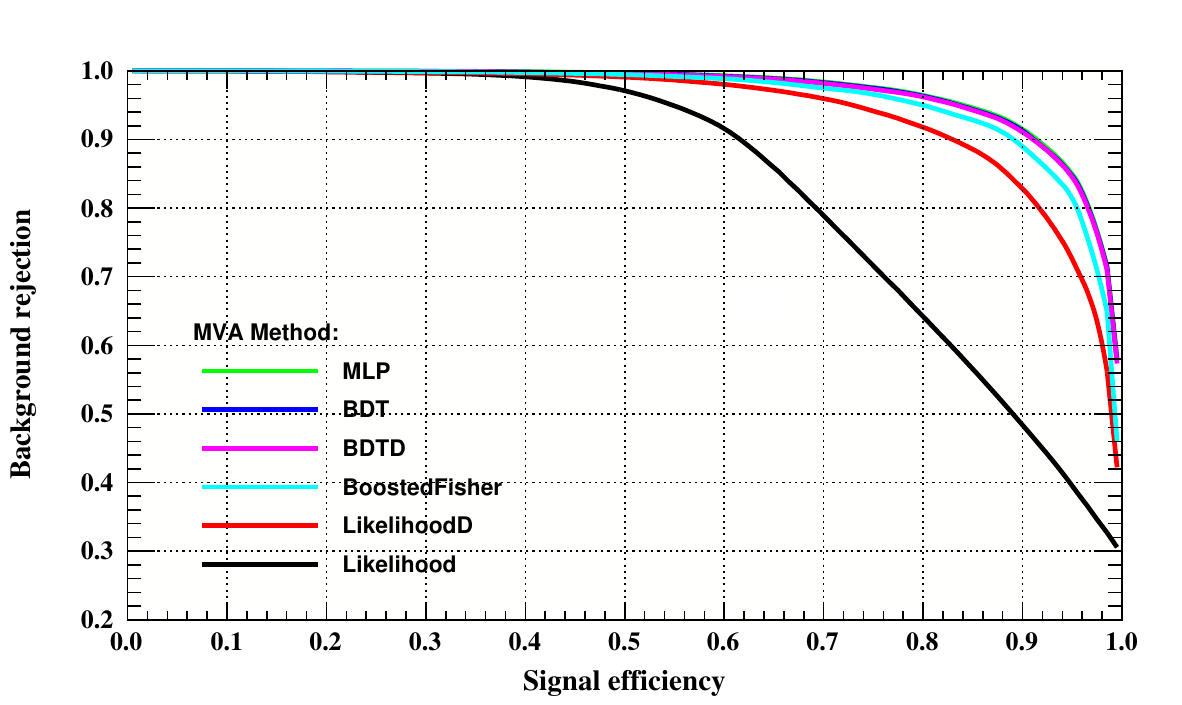}}
	\caption{Background rejection Vs signal efficiency for fully hadronic on the left and semi-leptonic on the right respectively.} 
	\label{ROC}
\end{figure}

%%%%%%%%%%%%%%%%%%%%%%%%%%%%%%%%%%%%%%%%%%%%%%%%%%%%%%%%
The curve of background rejection against signal efficiency can be a useful tool for accurately estimating a classifier's performance. High background rejection and high signal efficiency are required to achieve the greatest separation of noise from signal candidates. The expected separation power of a classifier improves with increasing area.
As we can see from Table \ref{auc of TMVA}, the area under the curve (AUC) for the third-generation leptoquarks (LQs) was evaluated by applying specific cut values to both decay modes. The AUC provides a quantitative measure of the performance of our signal identification methodology, particularly in distinguishing between the signal and the background events. From this table, the best classifier demonstrating the highest area under the curve (AUC) for the third-generation scalar leptoquark (sLQ) is the Multi-Layer Perceptron (MLP). The Receiver Operating Characteristic curves for both the hadronic and the semi-leptonic decay modes are shown in Fig. \ref{ROC}.
%%%%%%%%%%%%%%%%%%%%%%%%%%%%%%%%%%%%%%%%%%%%%%%%%%%%%%%%
%%%%%%%%%%%%%%%%%%%%%%%%%%%%%%%%%%%%%%%%%%%%%%%%%%%%%%%%
An analysis of the area under the curve (AUC) results that the semileptonic decay channel performs more proficiently than the fully hadronic channel in terms of discriminative power across all machine learning algorithms evaluated (MLP, BDT, BDTD, and Likelihood).
As seen in Figure \ref{ROC}, the MLP classifier shows the highest value of AUC for both decay modes, which implies that the MLP classifier is very appropriate for this particular classification problem. For instance, in the case of fully hadronic mode, this gives a value of $0.88$, similar behavior can be seen in semi-leptonic mode with a value of $0.97$ (by applying cuts). This may indicate that the design and training procedure of the MLP is successfully identifying the important characteristics and patterns in the data associated with the Leptoquark signal. It may also suggest that the MLP classifier is more resilient and broadly applicable to various datasets and situations about third-generation scalar Leptoquark detection applications. Other classifiers (BDT, BDTD and Likelihood) are showing prominent results with applying cuts, as shown in Table \ref{auc of TMVA}. This suggests that they can separate the positive and negative occurrences to a great degree by capturing the minute variations and patterns unique to the Leptoquark particles. Reaching maximum AUC values indicates that the MLP, BDT, BDTD, Likelihood, LikelihoodD and Boosted Fisher models are resilient to dataset variability and noise. They are less susceptible to being affected by random fluctuations or outliers and are capable of generalizing well to data that have not yet been observed. The ROC AUC score is a number between 0 and 1, with 1 representing excellent performance and $0.5$ indicating random guessing. While $0.7$ to $0.8$ is regarded as acceptable, $0.8$ to $0.9$ as good, and greater than $0.9$ as outstanding.
 %%%%%%%%%%%%%%%%%%%%%%%%%%%%%%%%%%%%%%%%%%%%%%%%%%%%%%%%
\begin{table}[h]
    \centering
    \begin{tabular}{||c|c|c||}\hline\hline
        \textbf{MVA Classifier} &\textbf{AUC (Semi-Leptonic )} &\textbf{AUC (Fully Hadronic)}\\\hline
        \textbf{MLP}&0.976& 0.886 \\\hline
         \textbf{BDT}&0.975& 0.884 \\\hline
         \textbf{BDTD}&0.975& 0.882 \\\hline
         \textbf{Likelihood}&0.956& 0.784\\\hline  
         \textbf{LikelihoodD}&0.956& 0.809\\\hline  
         \textbf{Boosted Fisher}&0.956& 0.864\\\hline 
    \end{tabular}
    \caption{The area under the Curve for $3^{rd}$ Gen LQs with cuts values for both of the decay mode. }
    \label{auc of TMVA}
\end{table}
%%%%%%%%%%%%%%%%%%%%%%%%%%%%%%%%%%%%%%%%%%%%%%%%%%%%%%%%
\section{Efficiency Cut of TMVA Methods}
%%%%%%%%%%%%%%%%%%%%%%%%%%%%%%%%%%%%%%%%%%%%%%%%%%%%%%%%
Each classifier is trained for 80000 signal events and 50000 background events. To check the total visibility of the signal process, signal significance is calculated for each classifier. 
%%%%%%%%%%%%%%%%%%%%%%%%%%%%%%%%%%
\begin{table}[h!]
\centering
\begin{tabular}{|c|m{6cm}|m{5cm}|}
\hline
\textbf{Decay Channel} & \textbf{Background Processes} & \textbf{Cross Section (pb)} \\
\hline
Fully Hadronic &
\(WWZ \), \( WW \), \( WZ \), \( ZZ \), \( t\bar{t} \), Single Top &
\begin{tabular}[t]{@{}l@{}}
\( WWZ: 0.01812 \) \\
\( WW: 27.67\) \\
\( WZ:  8.442 \) \\
\( ZZ: 2.536 \) \\
\( t\bar{t}: 220.3 \) \\
Single Top: 4.681 \\
\end{tabular} \\
\hline
Semi-Leptonic &
\( WWZ \), \( WW \), \( WZ \), \( ZZ \), \( t\bar{t}W \), \( t\bar{t}Z \) &
\begin{tabular}[t]{@{}l@{}}
\( WWZ: 0.006057 \) \\
\( WW: 9.212  \) \\
\( WZ: 0.009 \) \\
\( ZZ: 0.974 \) \\
\( t\bar{t}: 24.06 \) \\
\( t\bar{t}Z: 0.01426 \) \\
\end{tabular} \\
\hline
Signal (Hadronic) & Hadronically Decaying Signal Process & \(  0.0007652 \) \\
\hline
Signal (Semi-Leptonic) & Semileptonically Decaying Signal Process & \( 0.002118 \) \\
\hline
\end{tabular}
\caption{Background Processes and Cross Sections for Different Decay Channels}
\label{SM-B}
\end{table}
%%%%%%%%%%%%%%%%%%%%%%%%%%%%%%%%%%
 \begin{eqnarray}
        pp &\rightarrow & S_{1}\Bar{S_{1}},~S_{1}\rightarrow t\tau,~\Bar{S_{1}}\rightarrow \Bar{t}\Bar{\tau}, t\rightarrow b jj, \Bar{t} \rightarrow \Bar{b} jj
        \label{3}\\
pp &\rightarrow &S_{1}\Bar{S_{1}},~S_{1}\rightarrow t\tau,~\Bar{S_{1}}\rightarrow \Bar{t}\Bar{\tau}, t\rightarrow b w^{+},w^{+}\rightarrow l^{+} \nu l, \Bar{t} \rightarrow \Bar{b} , w^{-},w^{-}\rightarrow l^{-} \Tilde{\nu l}
\label{4}
\end{eqnarray}
Eq.\ref{3} represents a third-generation scalar leptoquark for fully hadronic decay, having two top quarks, two tau leptons, two b-jets, and four light jets in the final states.
Signal significance is calculated by using cuts as follows:

\begin{table}[h!]
\centering
\begin{tabular}{|l|c|c|}
\hline
\textbf{Cut Variable} & \textbf{Semileptonic Decay} & \textbf{Fully Hadronic Decay} \\
\hline
Jet transverse momentum ($p_T$) & $> 700\,\text{GeV}$ & $> 700\,\text{GeV}$ \\
\hline
Jet pseudorapidity ($|\eta|$) & $\leq 2.5$ & $\leq 2.5$ \\
\hline
Missing transverse energy ($E_T^{\text{miss}}$) & $> 500\,\text{GeV}$ & $> 250\,\text{GeV}$ \\
\hline
Number of jets ($N_{\text{jets}}$) & $\geq 4$ & $\geq 6$ \\
\hline
Mean squared delta R between jets ($\Delta R$) & $> 0.4$ & $> 0.4$ \\
\hline
Delta phi between jets ($\Delta \phi$) & $\leq 0.5$ & $\leq 0.5$ \\
\hline
Total Hadronic Transverse Energy (THT) & $> 1200\,\text{GeV}$ & $> 1500\,\text{GeV}$ \\
\hline
\end{tabular}
\caption{Selection cuts used for semileptonic and fully hadronic decay channels of third-generation scalar leptoquark (MVA).}
\end{table}

\begin{figure}[h]
 	\centering
   \subfigure[]{\label{fig:a}\includegraphics[width=80mm]{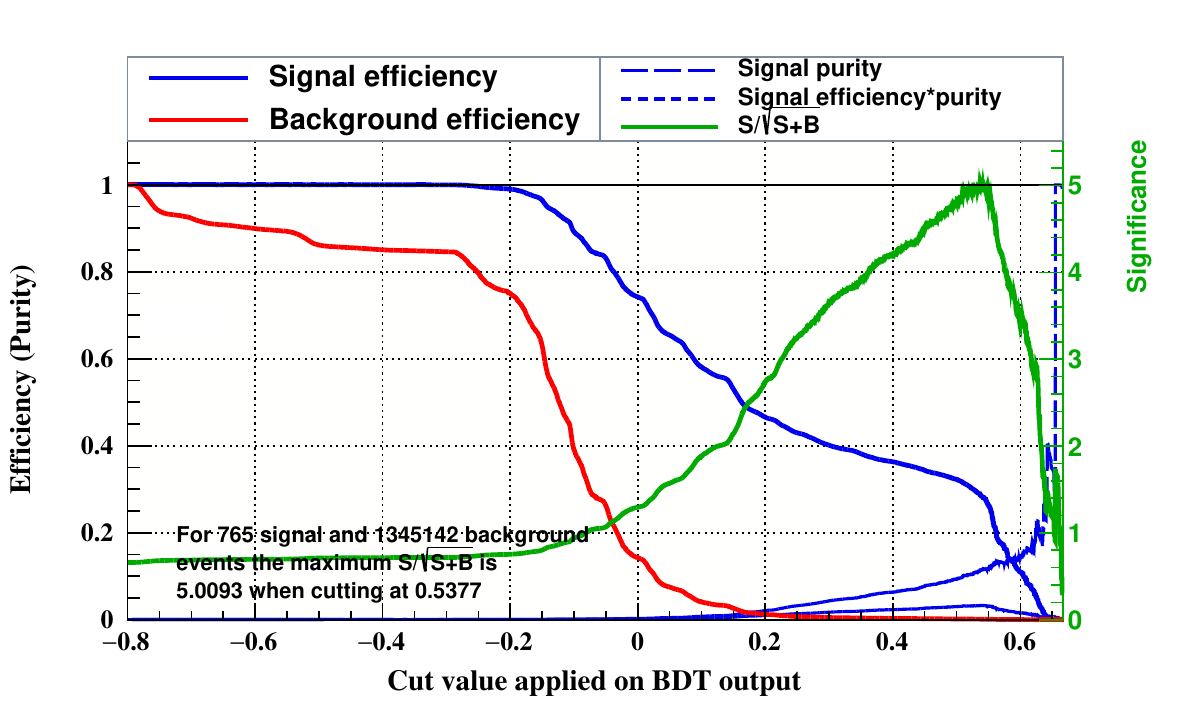}}
    \subfigure[]{\label{fig:b}\includegraphics[width=80mm]{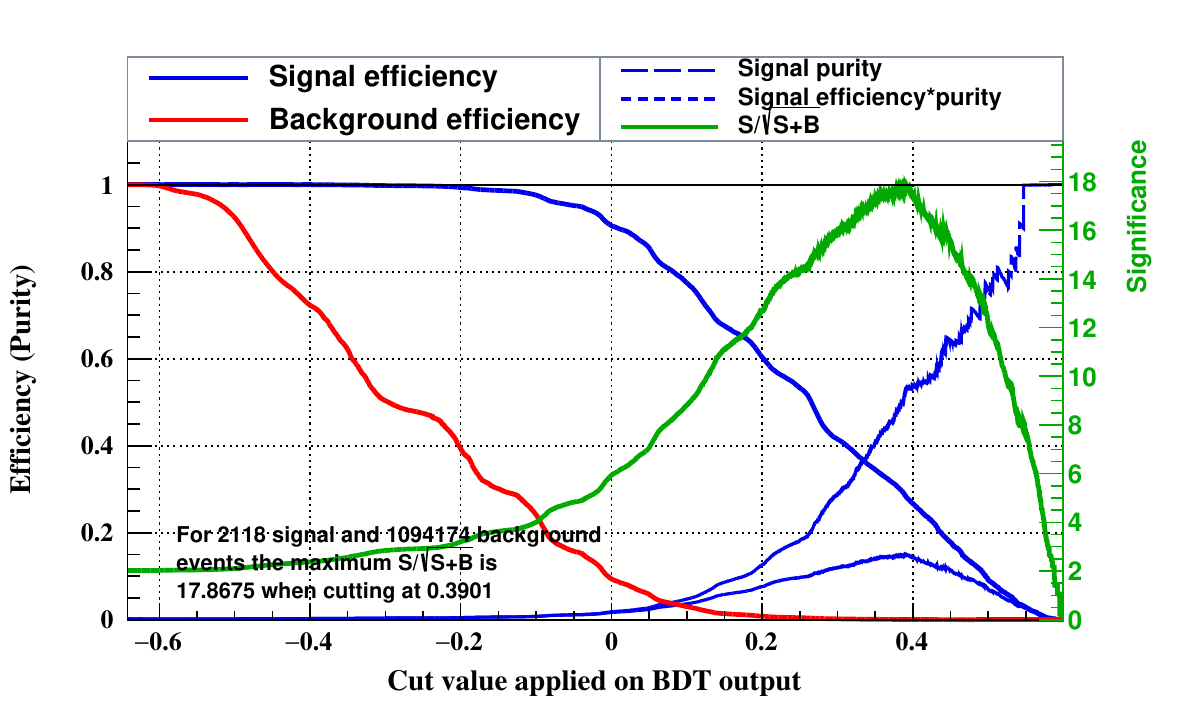}}
	\caption{BDT signal significance for fully hadronic on the  left and semi-leptonic on the right, respectively (with cuts)} 
	\label{HAD31}
\end{figure}

\begin{figure}[!ht]
	\centering
   \subfigure[]{\label{fig:a}\includegraphics[width=80mm]{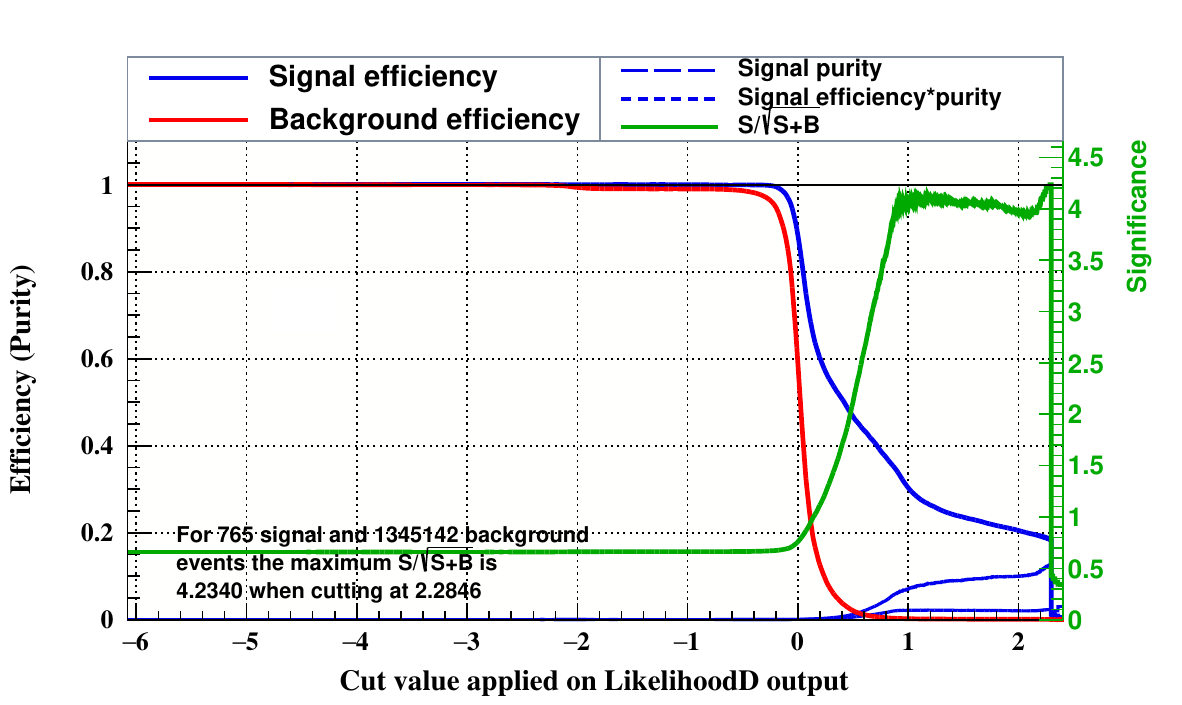}}
    \subfigure[]{\label{fig:b}\includegraphics[width=80mm]{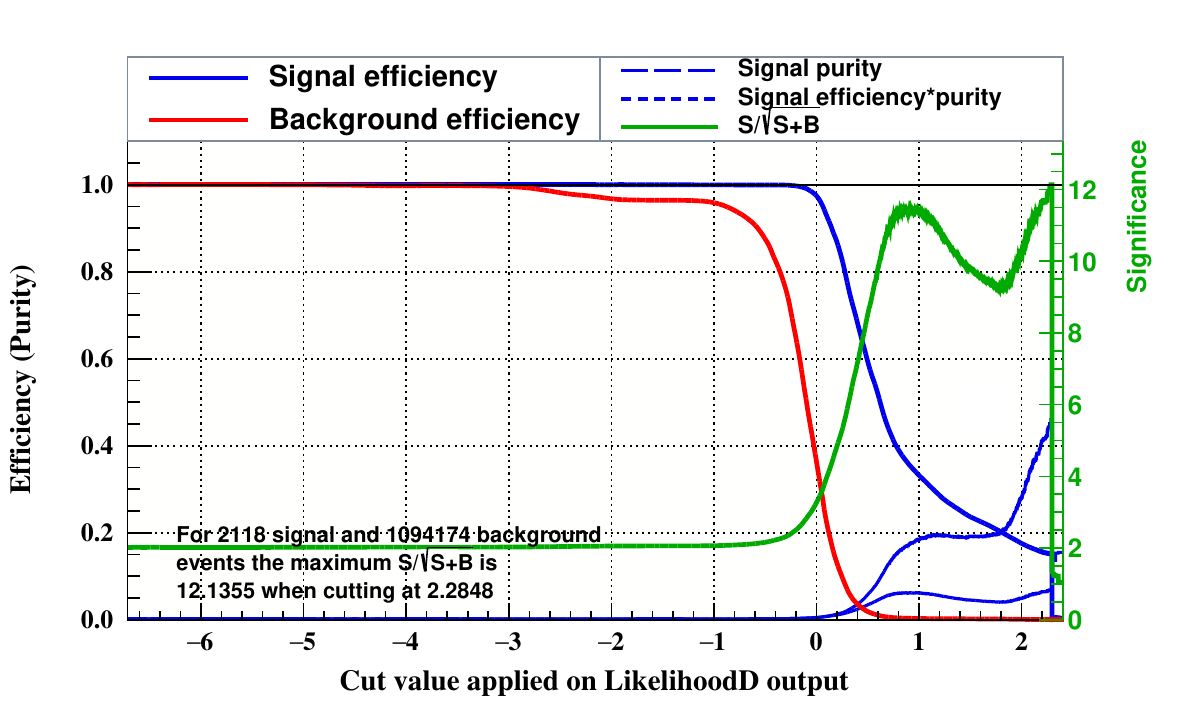}}
	\caption{LikelihoodD signal significance for fully hadronic on the left and semi-leptonic on the right, respectively (with cuts). } 
	\label{HAD32}
\end{figure}

\begin{figure}[!ht]
	\centering
  \subfigure[]{\label{fig:a}\includegraphics[width=80mm]{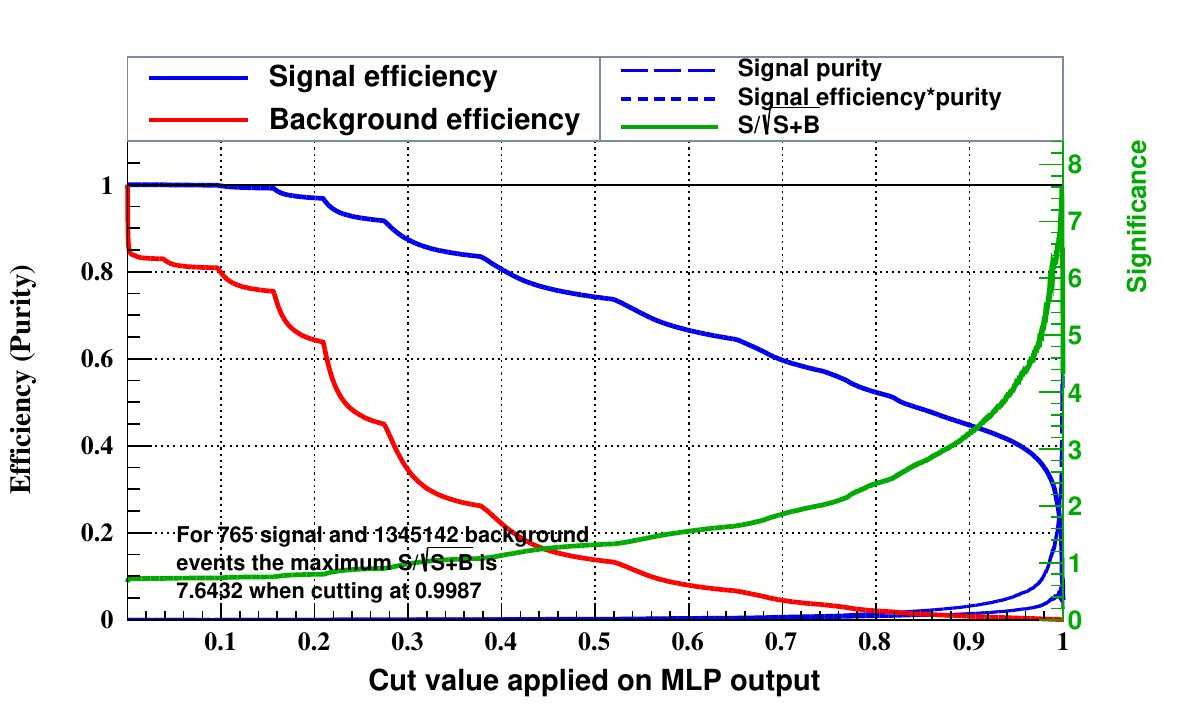}}
    \subfigure[]{\label{fig:b}\includegraphics[width=80mm]{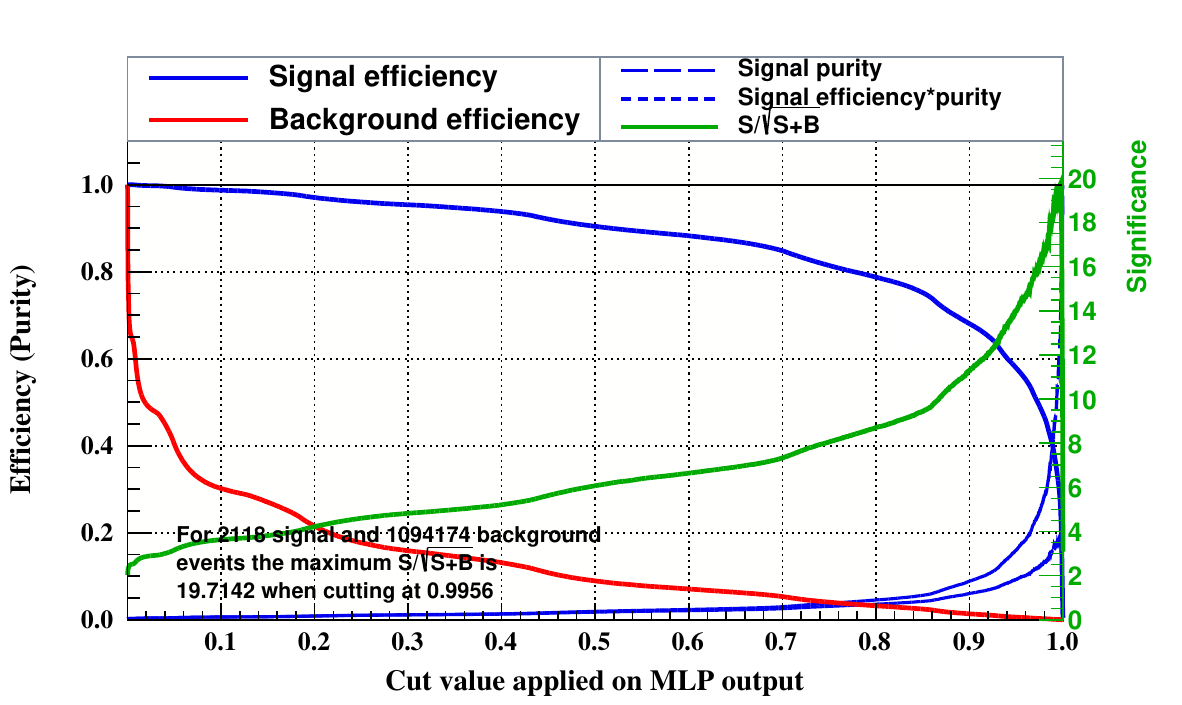}}
	\caption{MLP signal significance for fully hadronic on the left and semi-leptonic on the right, respectively (with cuts). }
	\label{HAD33}
\end{figure}

\begin{figure}[!ht]
	\centering
  \subfigure[]{\label{fig:a}\includegraphics[width=80mm]{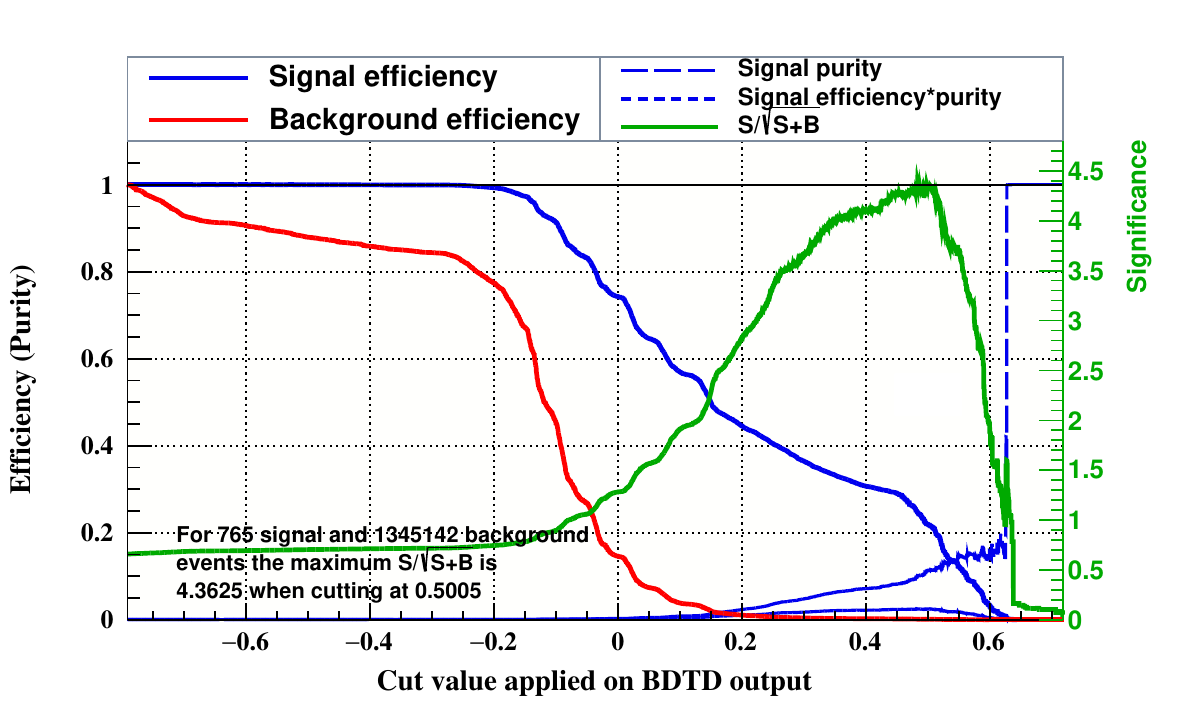}}
    \subfigure[]{\label{fig:b}\includegraphics[width=80mm]{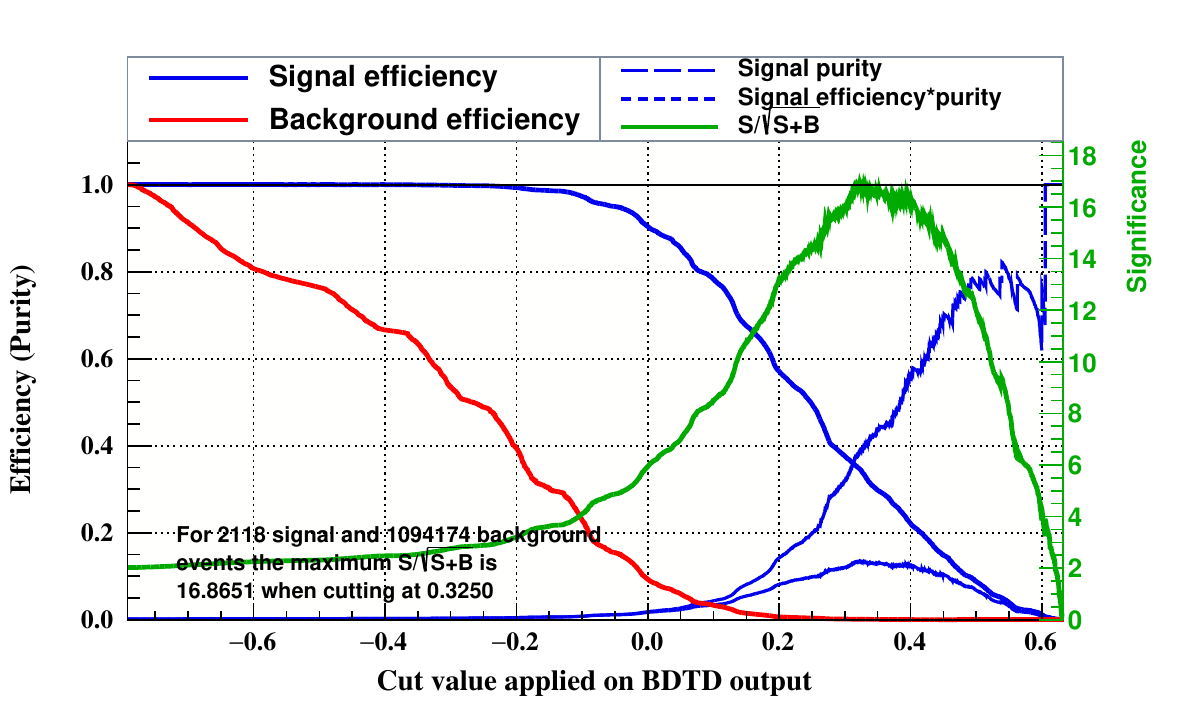}}
	\caption{BDTD signal significance for fully hadronic on the left and semi-leptonic on the right, respectively (with cuts). }
	\label{HAD34}
\end{figure}

\begin{figure}[!ht]
	\centering
  \subfigure[]{\label{fig:a}\includegraphics[width=80mm]{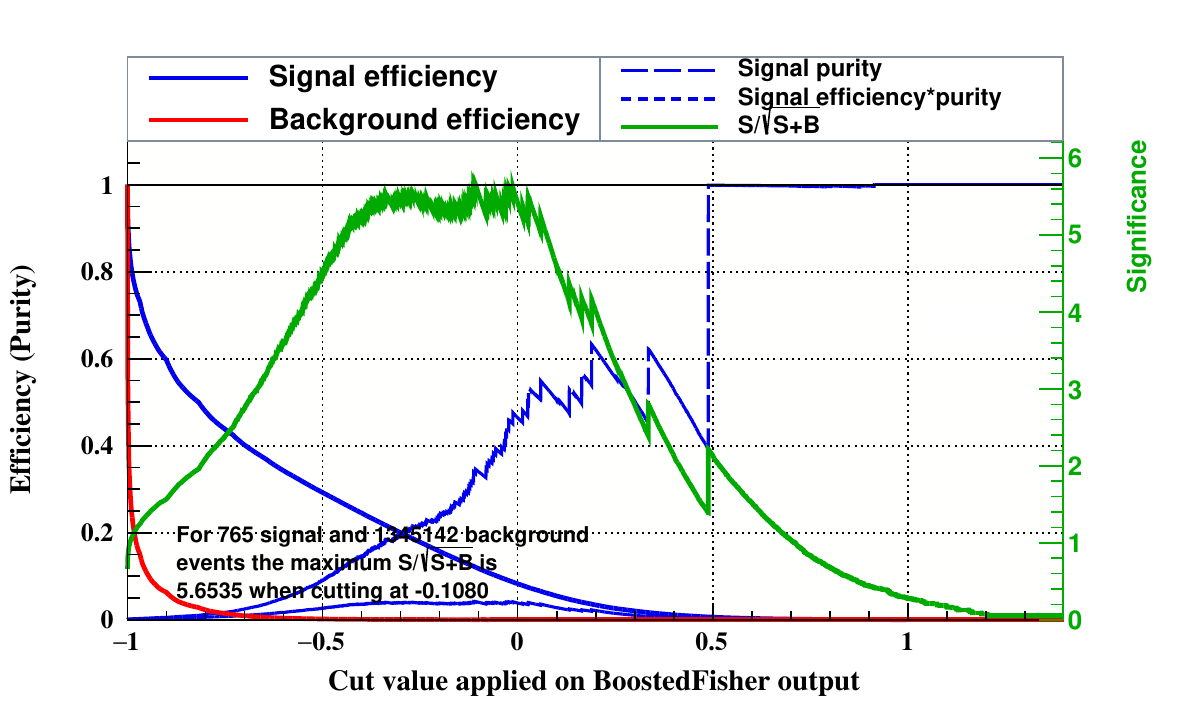}}
    \subfigure[]{\label{fig:b}\includegraphics[width=80mm]{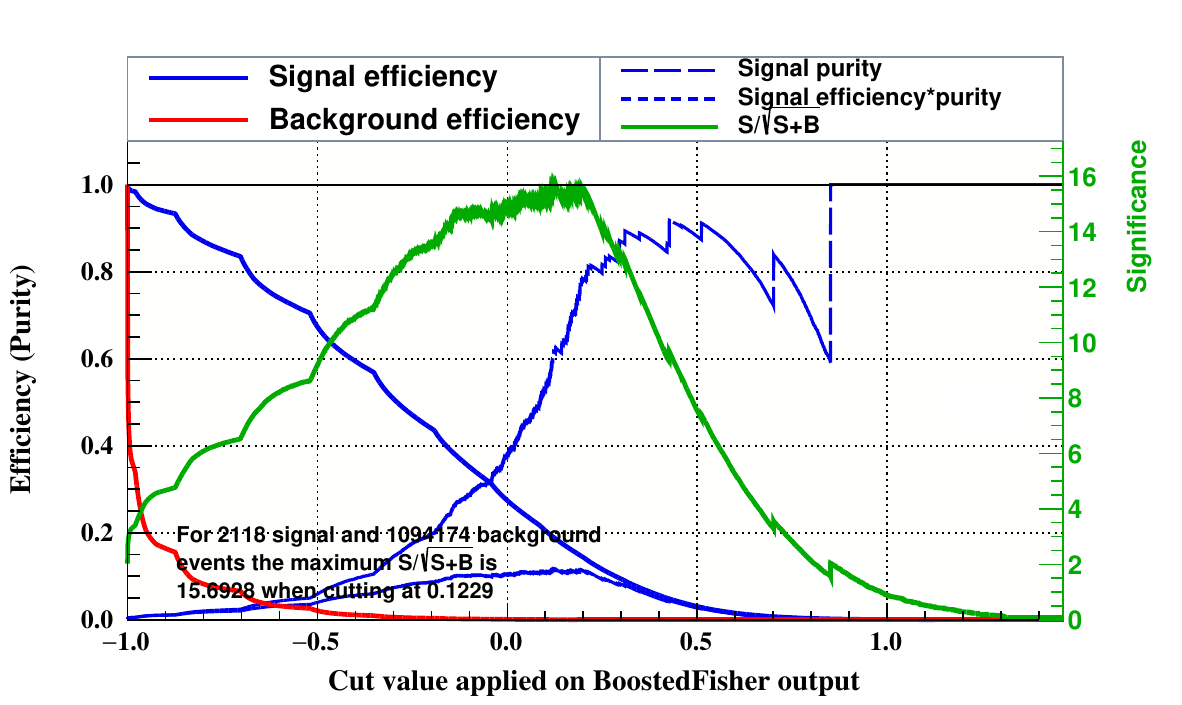}}
	\caption{The normalized Boosted Fisher statistics for background and signal events are displayed for both fully hadronic (left) and semi-leptonic (right) decay modes, where the BDT classifier explicitly separates the signal from the background. }
	\label{HADbf}
\end{figure}

%%%%%%%%%%%%%%%%%%%%%%%%%%%%%%%%%%
Figures. \ref{HAD31} - \ref{HADbf}, provide information on cut efficiencies and the optimal cut value for a given signal and background. These figures show the relationship between a classifier output and various metrics such as signal efficiency, background efficiency, signal purity, and significance separately for fully hadronic and semileptonic events. %As an example for the reader, Figure \ref{HAD35} demonstrates the response of one of the four normalized classifier output e.g., BDT for test and training samples for both fully hadronic and semi-leptonic decay channels. 
%%%%%%%%%%%%%%%%%%%%%%%%%%%%%%%%%%
%%%%%%%%%%%%%%%%%%%%%%%%%%%%%%%%%%
%%%%%%%%%%%%%%%%%%%%%%%%%%%%%%%%%%

%%%%%%%%%%%%%%%%%%%%%%%%%%%%%%%%%%
%%%%%%%%%%%%%%%%%%%%%%%%%%%%%%%%%%
%%%%%%%%%%%%%%%%%%%%%%%%%%%%%%%%%%
 \begin{figure}[!ht]
    \centering
    \includegraphics[width=9cm,height=6cm]{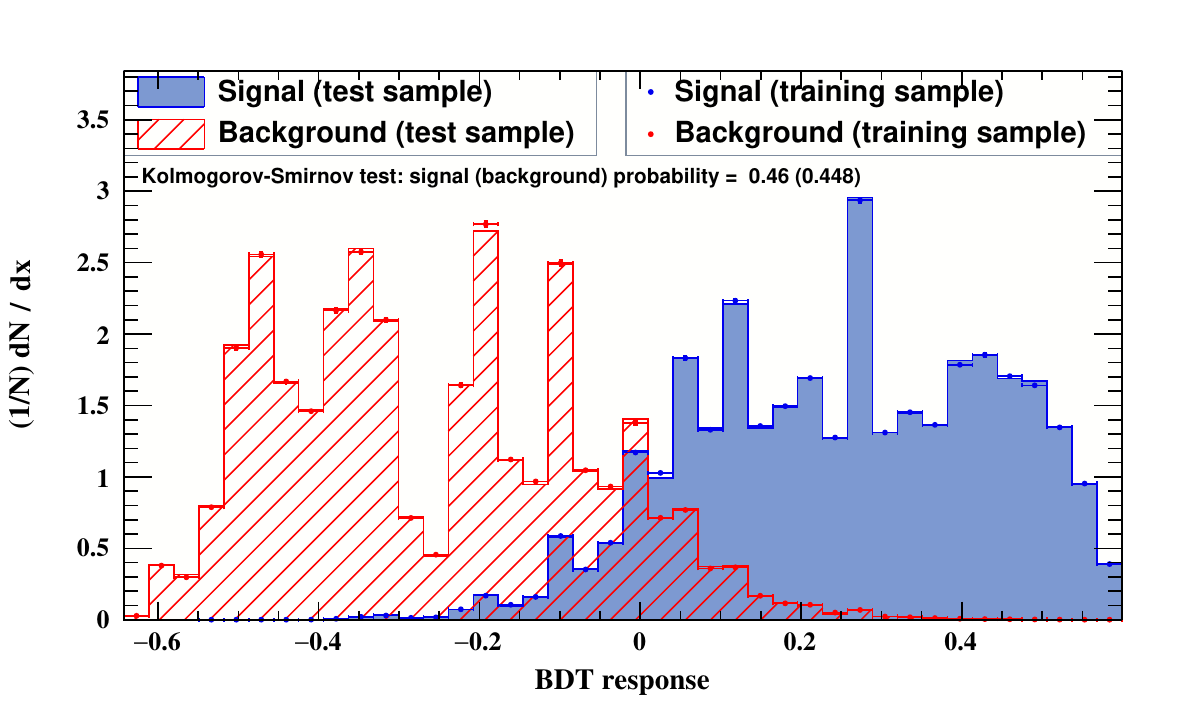}
    \caption{The normalized distribution of the BDT output for the training and testing samples for the background and signal classes.}
    \label{Res}
\end{figure}
%%%%%%%%%%%%%%%%%%%%%%%%%%%%%%%%%%
In Fig. \ref{Res} BDT classifier certainly distinguishes the signal from the background, with very little overlap in the middle region. 
 The classifier indicating maximum probability of signal $0.46$, while for background it is $0.44$. These value indicate an excellent arrangement among the training and test samples.   
The signal significance of the signal process is used in high-energy physics to quantify its separation. In the context of a third-generation scalar leptoquark (involving hadronic modes), the MLP as well as other classifiers yields the highest signal significance, along with the best signal efficiency and the background efficiency. This indicates that the classifier successfully and accurately differentiates between the signal (instances of the third-generation scalar leptoquark) and the background. This demonstrates that the classifier effectively detects the intended signal while minimizing background false positives. 
As it is clear that the MLP classifier achieves a signal significance of up to $7.64$ with applied cuts. Similarly, for the semi-leptonic decay mode, the MLP shows the maximum value of the signal significance when applying cuts have value of $19.7$, while for the BDT, BDTD, Likelihood, LikelihoodD and Boosted Fisher classifier significance reported in Table \ref{HC}.
 This finding suggests that the MLP design better captures the fundamental physics of the leptoquark, enabling it to achieve greater signal significance than other classifiers. The MLP's superior performance in this context may be attributed in part to its flexibility in modeling complex events and its capacity to understand nuanced correlations within the data.
%%%%%%%%%%%%%%%%%%%%%%%%%%%%%%%%%%%%%%%%%%%%%%%%%%%%%%%%%%%%%%%%%%%%%%%%%%%%%%%%%%%
\begin{table}[!ht]
\def\arraystretch{1.7}
   \centering
    \begin{tabular}{||c| c| c| c| c||}\hline
     \textbf{MVA  Classifier } &\textbf{Optimal-Cut}&$\mathbf{\dfrac{S}{\sqrt{S+B}}}$(fully hadronic) &$\mathbf{\dfrac{S}{\sqrt{S+B}}}$ (semileptonic) \\\hline
\textbf{MLP}&0.99&7.64& 19.71\\\hline
 \textbf{BDT}&0.53&5.0&17.86\\\hline 
\textbf{ BDTD}&0.50&4.36&16.86\\\hline   
\textbf{Likelihood}&2.13&6.2&13.35\\\hline
\textbf{LikelihoodD}&2.28&4.23&12.13\\\hline
\textbf{Boosted Fisher}&-0.10&5.65&15.69\\\hline
    \end{tabular}
   \caption{The optimal-cuts and the signal-background ratio with applying cuts at $\mathcal{L}_{\textit{int}}= 1000 fb^{-1}$ againts BP3 and BP4 respectively. }
    \label{HC}
\end{table}
 \begin{figure}[!ht]
    \centering
    \includegraphics[width=9cm,height=6cm]{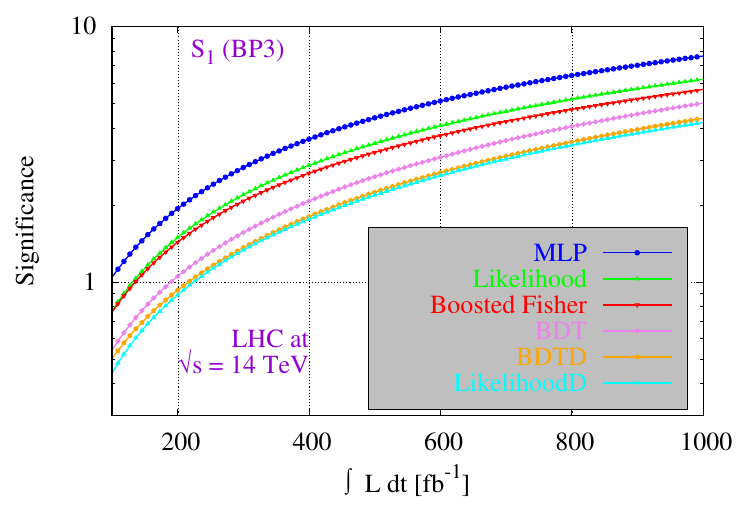}
    \caption{Signal significance vs integrated luminosity for $3^{rd}$ generation scalar leptoquark (BP3), considering machine learning algorithm.}
    \label{Sig vs lumi}
\end{figure}
%%%%%%%%%%%%%%%%%%%%%%%%%%%%%%%%%%%%%%%%%%%%%%%%%%%%%%%%%%%%%%%%%%%%
Fig. \ref{Sig vs lumi} shows the signal significance as a function of integrated luminosity, where we can conclude that the signal significance steadily increases and approaches 5$\sigma$ as the integrated luminosity reaches 1000$fb^{-1}$. This trend indicates that with a greater accumulation of data, the statistical fluctuations reduce, allowing for clearer distinctions between the signal and background noise. Consequently, the increased luminosity enhances the capability to confirm the signal's presence with high confidence, ultimately reinforcing the validity of the findings in particle physics research and allowing for more robust discoveries.
%%%%%%%%%%%%%%%%%%%%%%%%%%%%%%%%%%%%%%%%%%%%%%%%%%%%%%%%%%%%%%%%%%%%
\begin{table}[h!]
\centering

\begin{tabular}{||c|c|c|c||}\hline
%\toprule
\textbf{Benchmark Point} & \textbf{Tool} & \textbf{Decay Mode} & \textbf{Significance} \\\hline
%\midrule
BP1 & MA5 & Fully Hadronic   & 4.2 \\\hline
BP2 & MA5 & Semi- Leptonic  & 6.7 \\\hline
BP3 & MVA (Likelihood) & Fully Hadronic   & 6.2 \\\hline
BP4 & MVA (Likelihood) & Semi- Leptonic & 13.3 \\\hline
%\bottomrule
\end{tabular}
\caption{Comparison of significance for four benchmark points at $M_{LQ} = 1500$~GeV with an integrated luminosity of $\mathcal{L}_{\text{int}} = 1000~\mathrm{fb}^{-1}$.}
\label{comp}
\end{table}
%%%%%%%%%%%%%%%%%%%%%%%%%%%%%%%%%%%%%%%%%%%%%%%%%%%%%%%%%%%%%%%%%%%%
Table \ref{comp} shows a comparison between the signal significance \cite{MA5} based on the conventional cut-based method and multivariate analysis (MVA) based on neural networks for different BP's of scalar leptoquarks. This comparison indicates that MVA provides higher signal significance values compared to MadAnalysis5, such as at $m_{LQ}$ = 1500 GeV, where conventional cut-based method yields a significance of $4.2$ and, $6.7$ considering BP1 and BP2, while machine-learning based algorithm achieves values of $6.2$ and $13.3$  (Likelihood) for the fully hadronic and semi-leptonic modes against BP3 and BP4 respectively, applying the same selection cuts and at the same integrated luminosity.
Multiple factors can be considered concurrently by machine-learning based algorithm, enhancing their ability to distinguish between the signal and the background events. This capability is particularly valuable when working with complex signal distributions or overlapping backgrounds. Furthermore, machine-learning based algorithm can detect subtle variations that straightforward cut-based techniques might overlook by tailoring the selection criteria to the properties of the signal and the background distributions. 
%\begin{figure}[!ht]
 %   \centering
  %  \includegraphics[width=9cm,height=6cm]{pt00.eps}
   % \caption{Mass vs significance of $P_T^{j}$ by considering $3^{rd}~Gen~sLQ$ (hadronic decay mode) at three different int. luminosity.}
   % \label{significance}
%\end{figure}
%Figure \ref{significance} represents mass vs. significance at three different luminosities for third-generation scalar leptoquarks while considering its hadronic decay mode at $\sqrt{s}$ = 14 TeV. As we can see from Figure \ref{significance}, with increasing mass of leptoquark, the significance decreases accordingly. Such that at 1350 GeV and $1000~fb^{-1}$ the significance is $0.114$ and for 1650 TeV the significance drops to 0.0247. That might be because the phase space that is open to the decay products of the leptoquark decreases as its mass rises. As a result, there are fewer accessible decay channels, which lowers the number of events that could be noticed. 
%While at the higher luminosities 3000 $fb^{-1}$ and 1300 GeV the value of significance increases up to 0.197. Similarly, for 1650 TeV, the value of significance becomes 0.0428. A higher integrated luminosity causes the detector to record more collisions. The statistical significance of the detected signal over background fluctuations improved as a result of the larger number of signal occurrences caused by the enhanced statistics.

%%%%%%%%%%%%%%%%%%%%%%%%%%%%%%%%%%%%%%%%%%%%%%
\section{Conclusion}
%%%%%%%%%%%%%%%%%%%%%%%%%%%%%%%%%%%%%%%%%%%%%%
%{Here, we presented the findings of leptoquark searches at the TeV scale with neural networks of different decay modes to investigate the possibility of finding LQs, which are being seen by the LHC's ATLAS and CMS detectors. This article primarily focused on the semi-leptonic decay modes and the dominant disintegration of a scalar leptoquark into third-generation quarks and leptons in completely hadronic decays. The leptoquark pair production at $\sqrt{s}$ = 14 TeV produced well-defined cumulative cuts that were used to determine the final states. These benchmark points are listed in Table \ref{B-Table}.
%We accounted for mass Vs cross-section, energy vs. cross-section, $\lambda$ vs. $\sigma$ and $\lambda$ vs. $\Gamma_{partial}$ in our work.%  ]  

In our work, we have investigated the third-generation scalar leptoquark at the energy $\sqrt{s}=14~TeV$ with cut-based analysis, because scalar leptoquarks can explain the B-meson flavor anomalies, and in the framework of the Pati-Salam model, it explains the unification of the leptons and the quarks by extending the SM gauge group.\\ 
The cross-section of third-generation scalar leptoquark $14~TeV$ is $1.476~fb$ ( for $M_{LQ}=1350 ~GeV$) for hadronic decay. Similarly, for semileptonic decay, it is $4.08~fb$, and it is increases with the center of mass energy. With the comparison of mass and cross-section, we obtained that with higher masses, the cross-section of scalar leptoquark decreases due to less phase space, which involves fewer kinematic configurations. The decay width of scalar leptoquarks increases with smaller couplings, so heavier leptoquarks will result in larger values of decay width, which is crucial for the collider phenomenology. The kinematical analysis was then carried out, wherein the invariant mass of leptoquarks for various quark flavors, transverse momentum, and pseudorapidity ($\eta$) for integrated luminosities $300~fb^{-1}, 1000~fb^{-1} $ and $3000~fb^{-1}$ were reconstructed for the hadronic decay for BP1 and semi-leptonic decay for BP2.

In our machine-learning based algorithm, we have used different classifiers to discriminate the signals from background events with the machine learning algorithms. Applying cuts, the signal efficiency and the background rejection for the hadronic decay are larger than for the semi-leptonic decay. We can compare the AUC curve for the increase in hadronic decay to semi-leptonic decay for the MLP $10.2\%$; for BTD, it increases to a value $10.3\%$; for  BDTD,  it rises to a value $10.4\%$; for LikelihoodD, its value reaches a value $18.2\%$, and for the Boosted Fisher, the signal efficiency and the background rejection increase  $10.6\%$, respectively.

The output response of classifiers for the hadronic decay shows that the signal significance is much higher for MLP and Likelihood than for other classifiers. For the hadronic decay, the value of signal significance for MLP at the optimal cut of 0.99 and Likelihood at the optimal cut of 2.13 is 7.64 and 6.2, respectively. The output response of classifiers for the semi-leptonic decay shows that the signal significance is much higher for MLP and BDT than for other classifiers. For the semi-leptonic decay, the value of signal significance for MLP at the optimal cut of 0.39 and BDT at the optimal cut of 2.13 is 19.71 and 17.86, respectively. 

We have compared the significance produced by machine-learning based algorithm and conventional cut-based method at the scalar leptoquark mass $1500~GeV$. For the hadronic decay, the significance for conventional cut-based method is 4.2, and in machine-learning based algorithm it is 6.2 for the Likelihood classifier. For the semi-leptonic decay, the significance for conventional cut-based method is 6.7, and in machine-learning based algorithm it is 13.3 for the Likelihood classifier. For the other classifiers in MVA, the significance value is also higher, which shows how effective  MVA models are at suppressing the background events from the signal events. As a result, in our machine-learning based algorithm models, all classifiers produced promising results at  $1000~fb^{-1}$ integrated luminosity by training and testing the events.  

%%%%%%%%%%%%%%%%%%%%%%%%%%%%%%%%%%%%%%%%%%%%%%%%%%%%%%%%%%%%%%%%
\section{Acknowledgements}
%%%%%%%%%%%%%%%%%%%%%%%%%%%%%%%%%%%%%
We gratefully acknowledge support from the Simons Foundation and member institutions. The current submitted version of the manuscript is available on the arXiv pre-prints home page.

\section{Statements and Declarations}
\textbf{Funding} \\
The authors declare that no funds, grants, or other support were received during the preparation of this manuscript.\\
\textbf{Competing Interests}\\
The authors have no relevant financial or non-financial interests to disclose.\\

\textbf{Availability of data and materials}\\
Data sharing is not applicable to this article as no datasets were generated or analyzed during the current study.\\

%\begin{thebibliography}{9}

%\end{thebibliography}
%\printbibliography
\bibliographystyle{ieeetr}
\bibliography{reference.bib}
\end{document}